\documentstyle[preprint,tighten,floats,prd,aps]{revtex}
\begin{document}                

\def\dfrac#1#2{{\displaystyle {#1 \over #2}}}
\def\eq#1{Eq.~(\ref{eq:#1})}
\def\xfig#1{Fig.~\ref{fig:#1}}
\def\citex#1{$^{\cite{#1}}$}

\input{epsf}

\epsfverbosetrue
\def\fignonpertiii{
\begin{figure}[t]
\vspace{10pt}
\leavevmode
\centering
\epsfxsize=\hsize
  \epsfbox{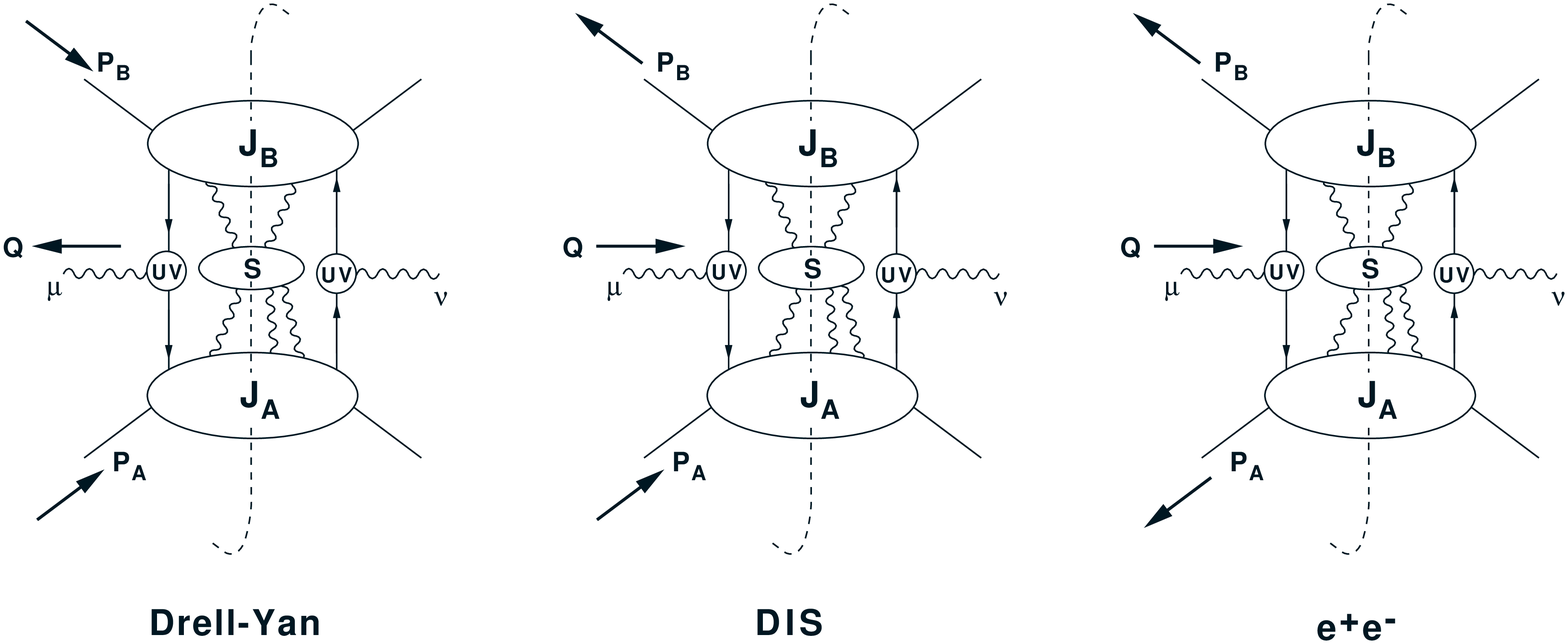}
      \caption{
Dominant integration regions leading to the
 non-perturbative contributions to
(a) Drell-Yan,
(b) DIS, and
(c) $e^+ e^-$.
 These three processes are related via a crossing symmetry.
 $J_A$ and $J_B$ represent the jet subgraphs associated with the collinear
partons from hadron $A(P_A)$ and $B(P_B)$, respectively.
 $S$ represents the subgraph of soft gluons and quarks which are connected
to the rest of the process by soft gluons (but not soft quarks).
 The double-logarithms  arise from  $J_A$ and $J_B$.
}
   \label{fig:nonpertiii}
\end{figure}
}
\def\fighadronframe{
\begin{figure}[t]
\vspace{10pt}
\leavevmode
\centering
\epsfxsize=3in
  \epsfbox{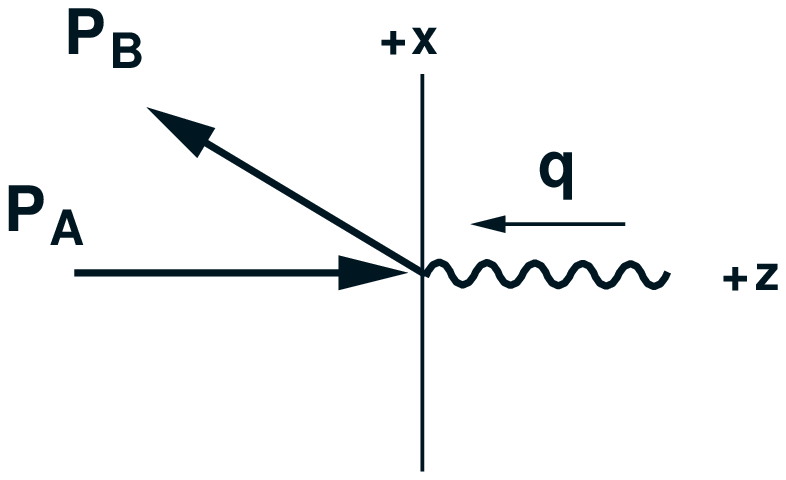}
      \caption{
The hadron frame.
The initial hadron $P_A$ lies along the positive $z$-axis, and
the vector boson $q$ lies along the negative $z$-axis.
The next-to-leading order QCD corrections can give the final state
hadron $P_B$ a non-zero $x$-component.
}
   \label{fig:hadframe}
\end{figure}
}
\def\figkini{
\begin{figure}[t]
\vspace{10pt}
\leavevmode
\centering
\epsfxsize=6in
  \epsfbox{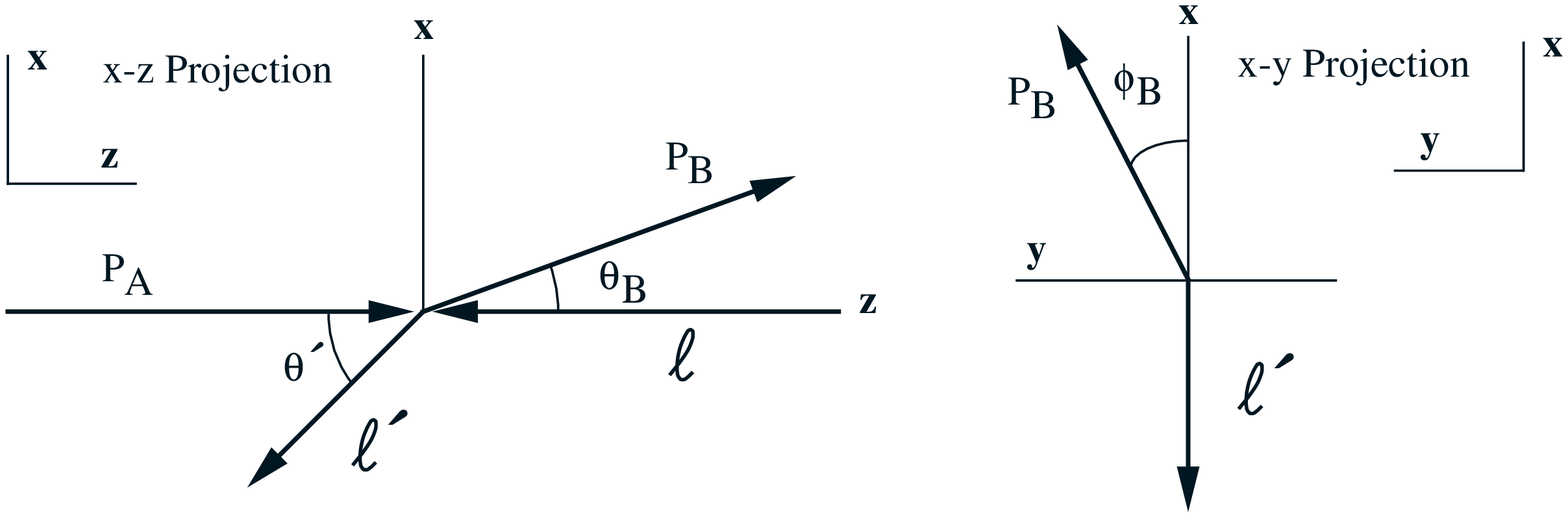}
      \caption{
The HERA lab frame for
the process:
$e^-(\ell) + A(P_A) \to e^-(\ell') + B(P_B) + X$.
The final state leption $e^-(\ell')$ lies in the
$x$-$z$--plane, and the final state hadron $B(P_B)$
has a non-zero $y$-component  if $\phi_B$ is non-zero.
}
   \label{fig:figkin1}
\end{figure}
}
\def\figqtphi{
\begin{figure}[t]
\vspace{10pt}
\leavevmode
\centering
\epsfxsize=4in
  \epsfbox{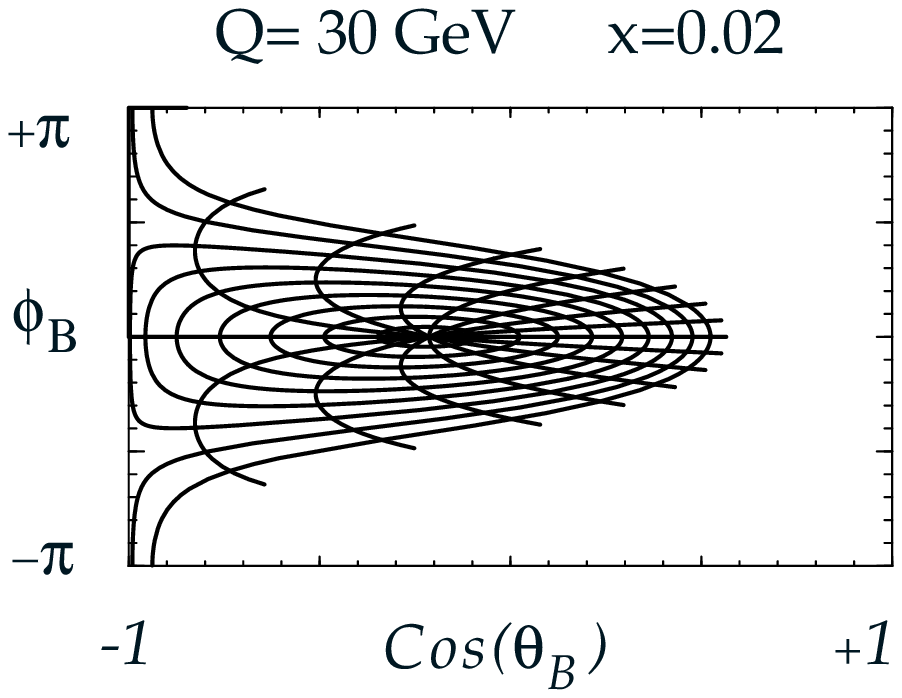}
      \caption{Contours in $\phi_B$ and $\cos(\theta_B)$ for $Q=30\, {{\rm
GeV}}$
       and $x=0.02$.
       The circular rings are contours of constant ${q_T}$ in steps of 3~{{\rm
GeV}}, and
       the radial arcs are contours of constant $\phi$ in steps of $\pi/8$.
}
  \label{fig:phietaii}
\end{figure}
}
\def\figdiagi{
\begin{figure}[t]
\vspace{10pt}
\leavevmode
\centering
\epsfxsize=4in
  \epsfbox{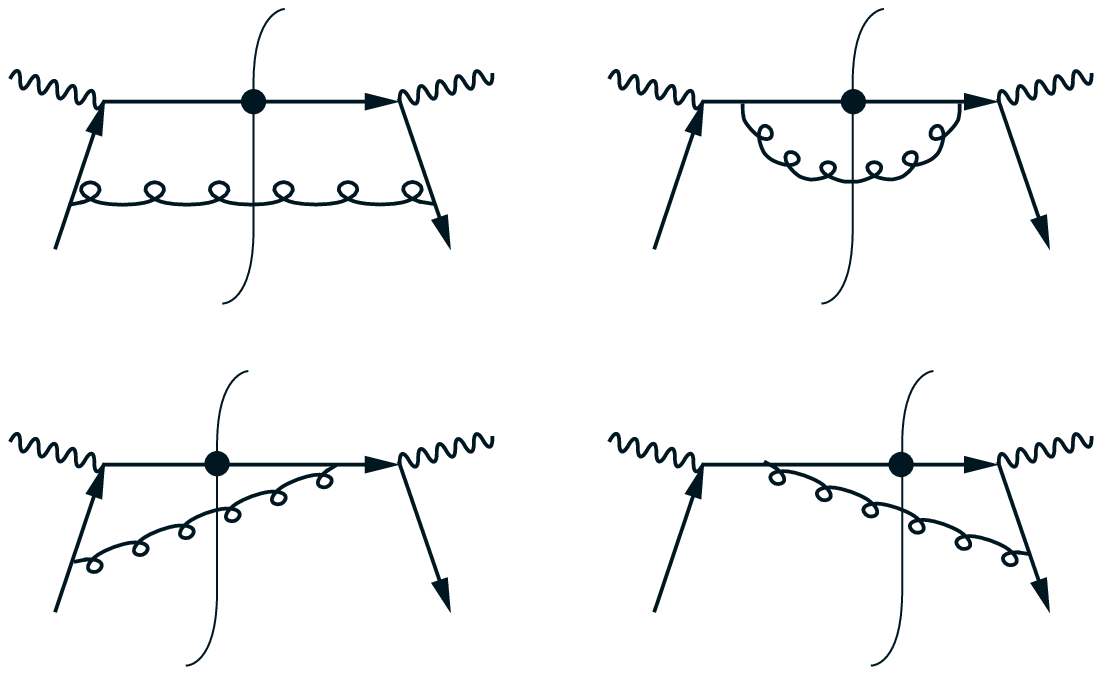}
      \caption{
Feynman diagrams for quark initiated process with a quark jet observed.
The observed parton is the upper line, indicated with a dot.
   }
   \label{fig:diai}
\end{figure}
}
\def\figdiagii{
\begin{figure}[t]
\vspace{10pt}
\leavevmode
\centering
\epsfxsize=4in
  \epsfbox{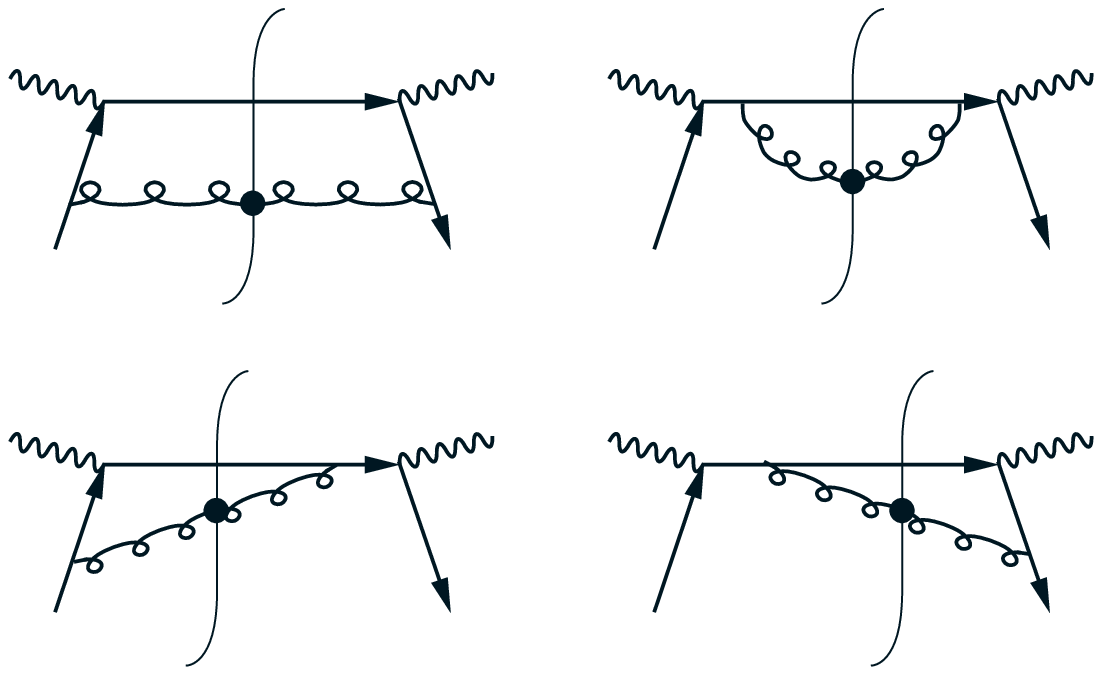}
      \caption{
Feynman diagrams for quark initiated process with a gluon jet observed.
The observed parton is the lower line, indicated with a dot.
   }
   \label{fig:diaii}
\end{figure}
}
\def\figdiagiii{
\begin{figure}[t]
\vspace{10pt}
\leavevmode
\centering
\epsfxsize=4in
  \epsfbox{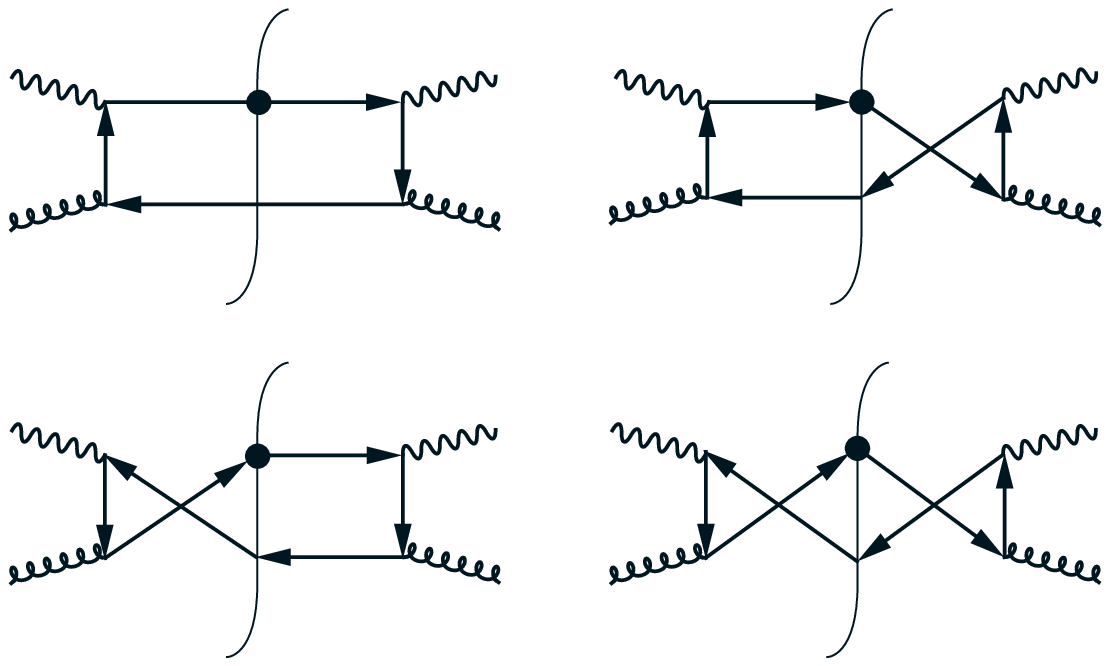}
      \caption{
Feynman diagrams for gluon initiated process with a quark jet observed.
The observed parton is the upper line, indicated with a dot.
   }
   \label{fig:diaiii}
\end{figure}
}
\def\figtempi{
\begin{figure}[t]
\vspace{10pt}
\leavevmode
\centering
\epsfxsize=4in
  \epsfbox{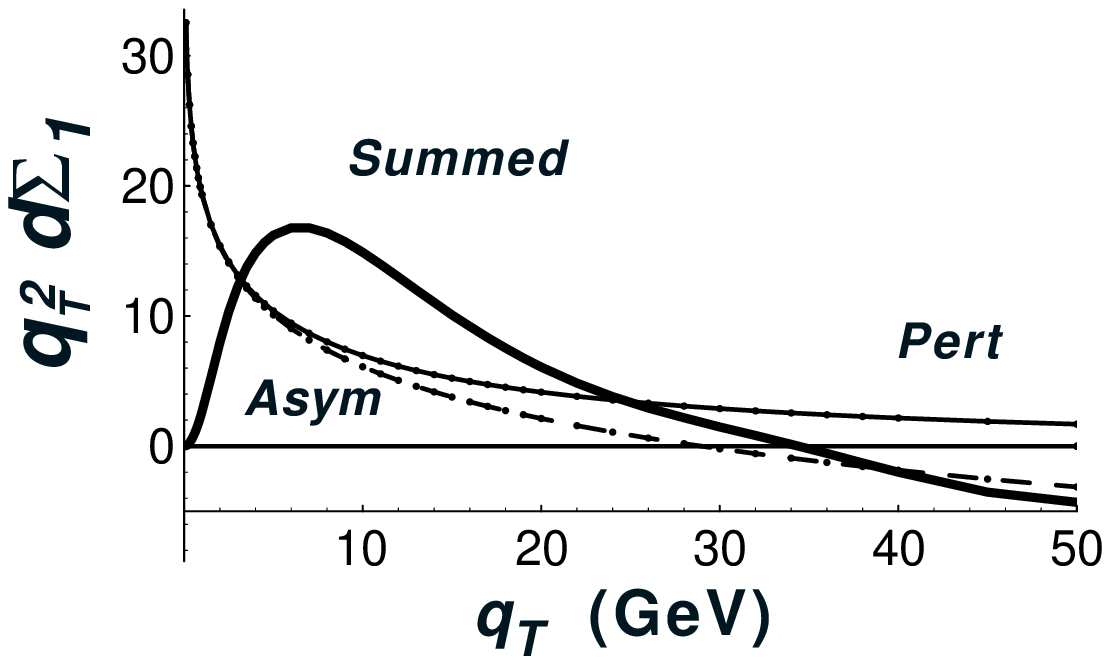}
      \caption{
The contributions to the energy distribution function
${q_T^2} \, d\Sigma_1/(dx\, dQ^2 \, d{q_T^2}\, d\phi)$
as a function of ${q_T}$,
for $Q=100\, {{\rm GeV}}$, $x=0.3$.
(Recall, $d\Sigma_1$ and $d\Sigma_6$ are independent of $\phi$.)
Perturbative (thin), asymptotic (dashed), and summed (thick).
Note how the perturbative and asymptotic cancel as ${q_T}\rightarrow 0$.
For ${q_T}\rightarrow Q$, the asymptotic and summed cancel to leading order
only.
 (A zero reference line is indicated.)
$d\Sigma_1$ is in units of ${{\rm GeV}}^{-5}$,
and is multiplied by $10^{9}$ for clarity of the plot.
   }
   \label{fig:tempi}
\end{figure}
}
\def\figtempii{
\begin{figure}[t]
\vspace{10pt}
\leavevmode
\centering
\epsfxsize=\hsize
  \epsfbox{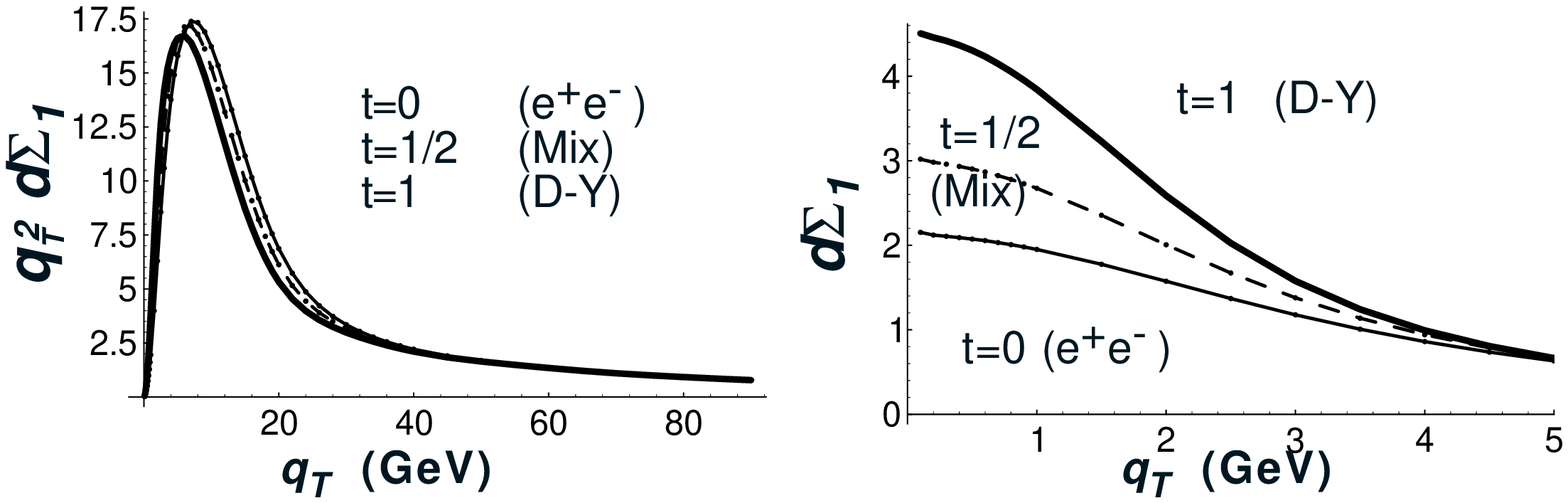}
      \caption{
The total contribution  to the energy distribution function
$d\Sigma_1/(dx\, dQ^2 \, d{q_T^2}\, d\phi)$ as a function of ${q_T}$
for different choices of the non-perturbative function, $S_{NP}(b)$,
for $Q=100\, {{\rm GeV}}$, $x=0.3$.
 Fig.~(a) has an extra factor of ${q_T^2}$  to make the plot more legible.
 Fig.~(b) demonstrates that the
summed contribution has a finite limit as ${q_T}\to 0$.
 We vary the $t$-parameter from $t=1$ (thick) corresponding to the Drell-Yan
case,
 to $t=1/2$ (dashed) corresponding to the mixed case,
 to $t=0$ (thin) corresponding to the $e^+ e^-$ case.
For ${q_T}\rightarrow Q$, we use the function ${\cal T}({q_T}/Q)$ with $\rho=5$
to smoothly switch between  large and small ${q_T}$.
$d\Sigma_1$ is in units of ${{\rm GeV}}^{-5}$,
and is multiplied by $10^{9}$ for clarity of the plot.
   }
   \label{fig:tempii}
\end{figure}
}
\def\figtempiii{
\begin{figure}[t]
\vspace{10pt}
\leavevmode
\centering
\epsfxsize=4in
  \epsfbox{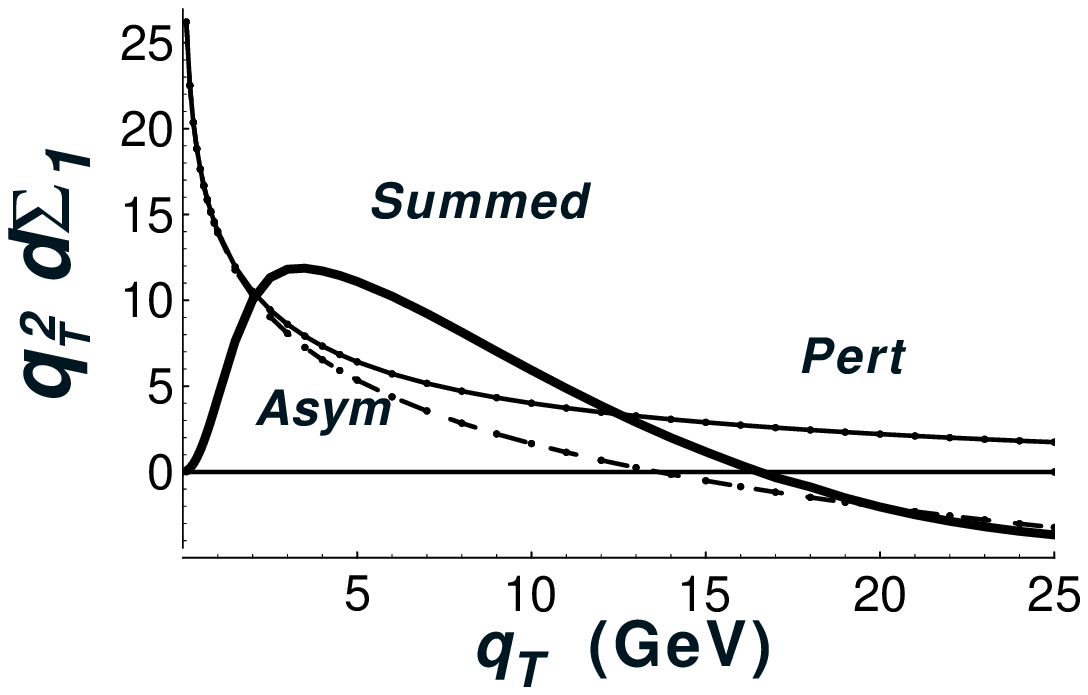}
      \caption{
The contributions to the energy distribution function
${q_T^2} \, d\Sigma_1/(dx\, dQ^2 \, d{q_T^2}\, d\phi)$ as a function of
${q_T}$,
for $Q=30\, {{\rm GeV}}$, $x=0.1$.
(Recall, $d\Sigma_1$ and $d\Sigma_6$ are independent of $\phi$.)
Perturbative (thin), asymptotic (dashed), and summed (thick).
Note how the perturbative and asymptotic cancel as ${q_T}\rightarrow 0$.
For ${q_T}\rightarrow Q$, the asymptotic and summed cancel  to leading
order only.
 (A zero reference line is indicated.)
$d\Sigma_1$ is in units of ${{\rm GeV}}^{-5}$,
and is multiplied by $10^{6}$ for clarity of the plot.
   }
   \label{fig:tempiii}
\end{figure}
}
\def\figtempiv{
\begin{figure}[t]
\vspace{10pt}
\leavevmode
\centering
\epsfxsize=\hsize
  \epsfbox{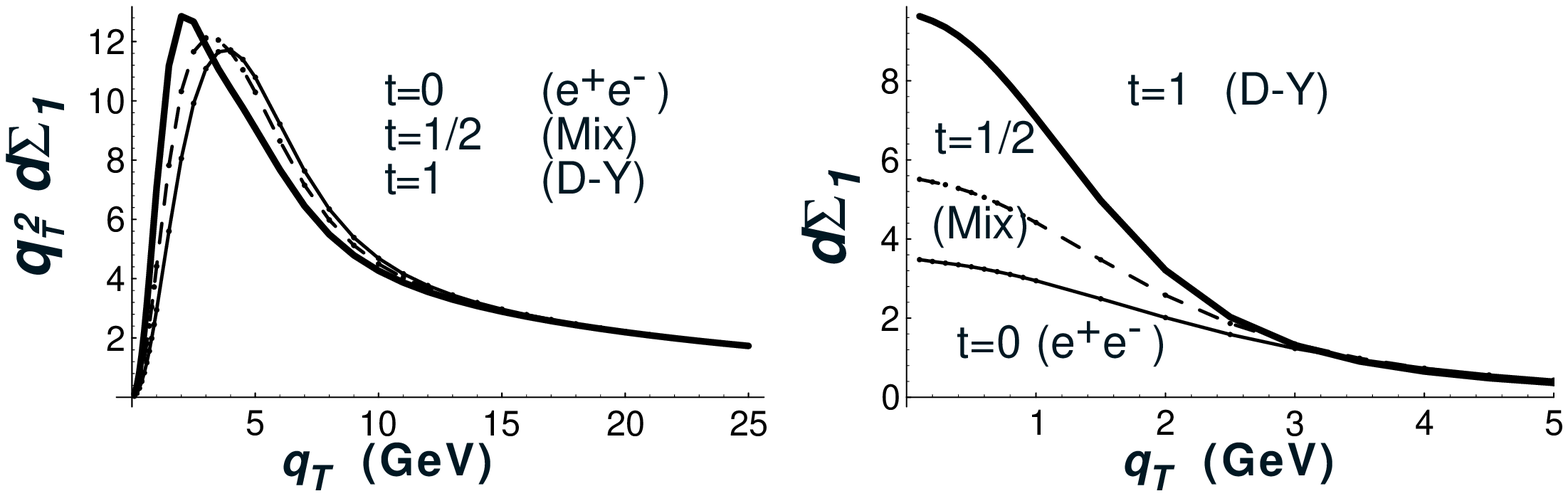}
      \caption{
The total contribution  to the energy distribution function
$d\Sigma_1/(dx\, dQ^2 \, d{q_T^2}\, d\phi)$ as a function of ${q_T}$
(scaled by $10^{9}$),
for different choices of the non-perturbative function, $S_{NP}(b)$,
for $Q=30\, {{\rm GeV}}$, $x=0.1$.
 Fig.~(a) has an extra factor of ${q_T^2}$  to make the plot more legible.
 Fig.~(b) demonstrates that the
summed contribution has a finite limit as ${q_T}\to 0$.
 We vary the $t$-parameter from $t=1$ (thick) corresponding to the Drell-Yan
case,
 to $t=1/2$ (dashed) corresponding to the mixed case,
 to $t=0$ (thin) corresponding to the $e^+ e^-$ case.
For ${q_T}\rightarrow Q$, we use the function ${\cal T}({q_T}/Q)$ with $\rho=5$
to smoothly switch between  large and small ${q_T}$.
$d\Sigma_1$ is in units of ${{\rm GeV}}^{-5}$,
and is multiplied by $10^{6}$ for clarity of the plot.
   }
   \label{fig:tempiv}
\end{figure}
}
\def\figtempv{
\begin{figure}[t]
\vspace{10pt}
\leavevmode
\centering
\epsfxsize=\hsize
  \epsfbox{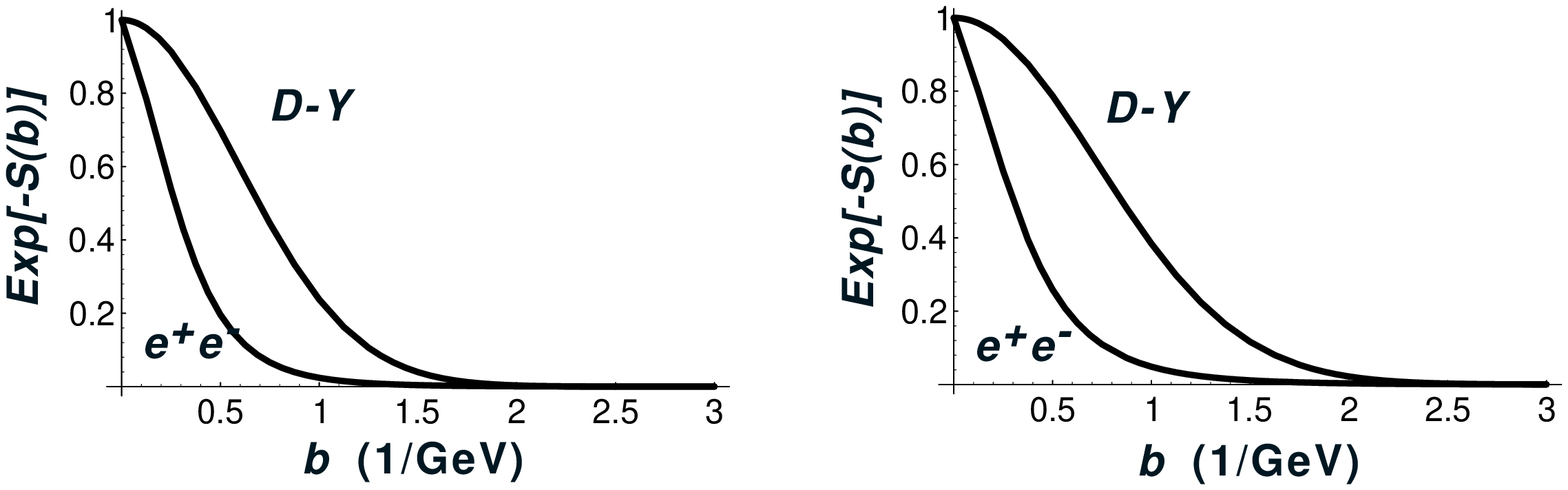}
      \caption{
      Comparison of the non-perturbative function $e^{-S_{NP}(b)}$
      {\it vs.} $b$ for the
     Drell-Yan (Davies, Webber, and Stirling\protect\cite{DWS}) [upper line],
     and  $e^+ e^-$ (Collins and Soper\protect\cite{CS})  [lower line],
    for $Q=30\, {{\rm GeV}}$ [Fig.~(a)], and $Q=100\, {{\rm GeV}}$ [Fig.~(b)].
    }
   \label{fig:tempv}
\end{figure}
}
\def\figtempvi{
\begin{figure}[t]
\vspace{10pt}
\leavevmode
\centering
\epsfxsize=\hsize
  \epsfbox{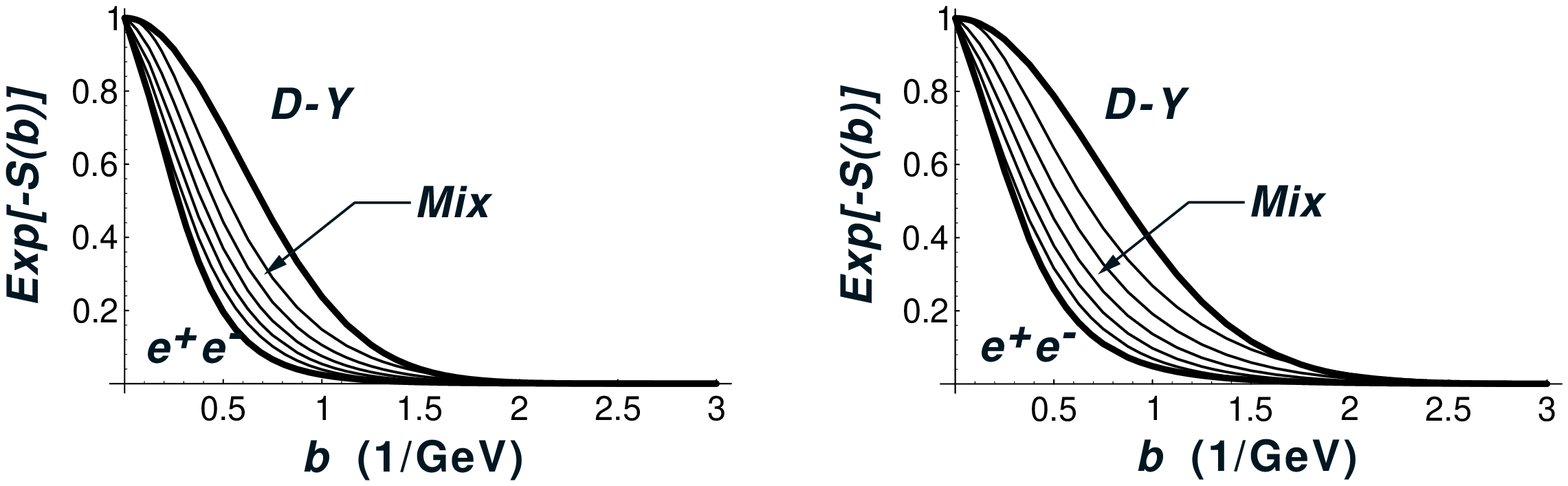}
      \caption{
      Interpolation of the non-perturbative function $e^{-S_{NP}(b)}$
      {\it vs.} $b$ as a function of the $t$-parameter,
      $\{t=0,1/4,1/2,3/4,1\}$,
      for $Q=30\, {{\rm GeV}}$ [Fig.~(a)], and $Q=100\, {{\rm GeV}}$
[Fig.~(b)].
      Note the variation of $e^{-S_{NP}(b)}$ as $t$ ranges over $[0,1]$
      is narrower than the full range between the   Drell-Yan
     and  $e^+ e^-$ case, ({\it cf.}, Sec.~\protect\ref{NONPERT}).
    }
   \label{fig:tempvi}
\end{figure}
}
\def\figtempvia{
\begin{figure}[t]
\vspace{10pt}
\leavevmode
\centering
\epsfxsize=\hsize
  \epsfbox{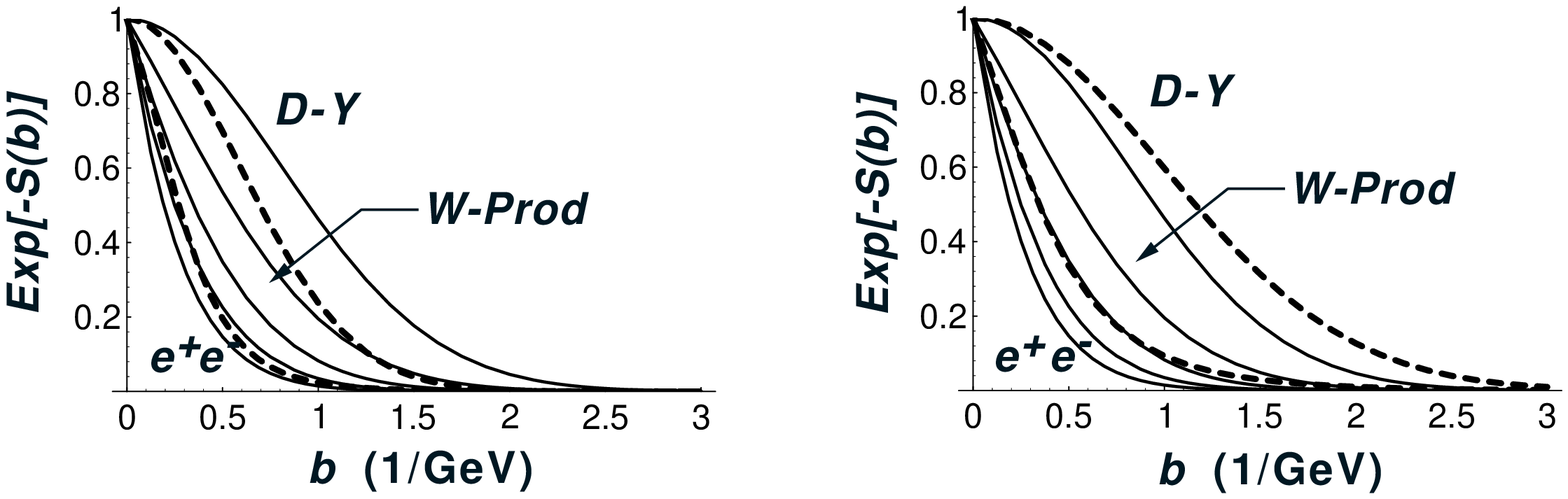}
      \caption{
      Comparison of the non-perturbative function $e^{-S_{NP}(b)}$
      {\it vs.} $b$ for the case of
     Drell-Yan (Davies, Webber, and Stirling\protect\cite{DWS}) [upper dashed
line],
     $e^+ e^-$ (Collins and Soper\protect\cite{CS})  [lower dashed line],
     and $W$-production (Ladinsky and Yuan\protect\cite{yuanladinsky}) [5 solid
lines].
    Fig.~(a) is for $Q=10\,  {{\rm GeV}}$ (left),  and
    Fig.~(b) is for $Q=100\, {{\rm GeV}}$ (right).
     The fits to $W$-production  include an extra parameter $\tau=\hat{s}/s$;
     we allow $\tau$ to range over the values
     $\tau=\{10^{-3}, 10^{-2.75}, 10^{-2.5}, 10^{-2.25}, 10^{-2.0} \}$ where
     $\tau=10^{-2}$ is the  upper curve, and
     $\tau=10^{-3}$ is the lower  curve in $b$-space.
}
   \label{fig:tempvia}
\end{figure}
}
\def\figtempvii{
\begin{figure}[t]
\vspace{10pt}
\leavevmode
\centering
\epsfxsize=4in
  \epsfbox{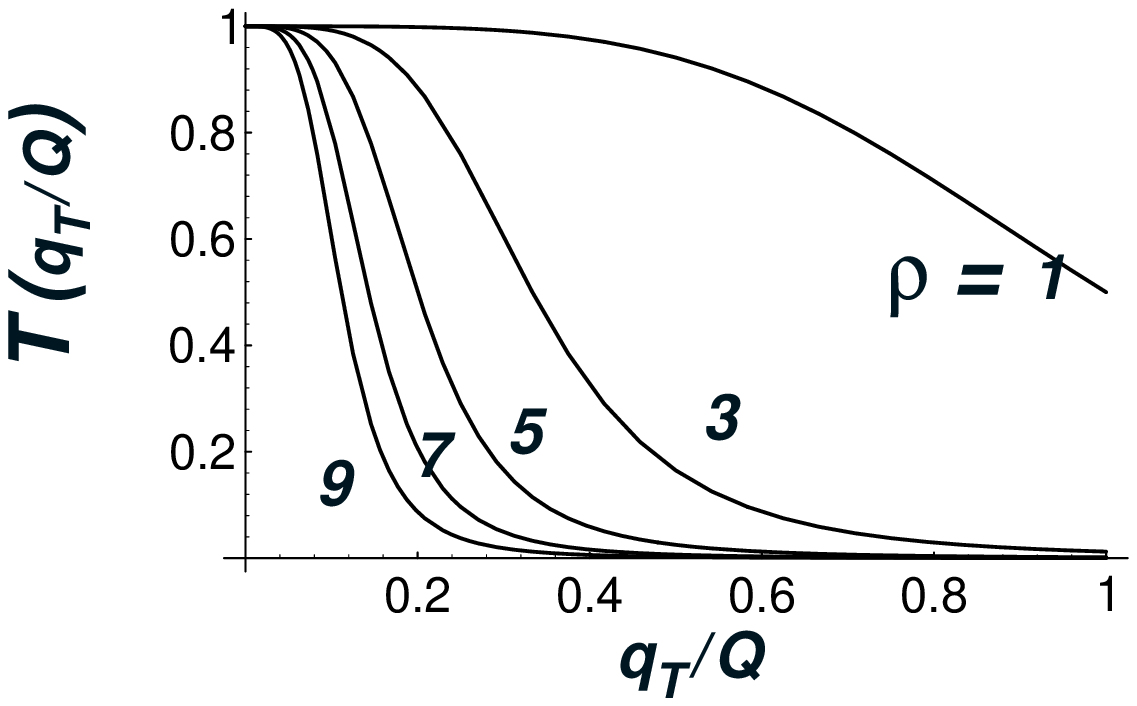}
      \caption{
     The matching function ${\cal T}({q_T}/Q)$
     {\it vs.} $({q_T}/Q)$ for
     $\rho=\{1,3,5,7,9\}$.
     $\rho=1$ is the top curve, and
     $\rho=9$ is the bottom curve.
    }
   \label{fig:tempvii}
\end{figure}
}



\preprint{\vbox{
\hbox{hep-ph/9511311}
\hbox{CTEQ-409}
}  }

\title{Semi-Inclusive Deeply Inelastic Scattering at Small ${q_T}$}

\author{ Ruibin Meng,{${}^{a}$}
    Fredrick I. Olness{${}^{b}$},
   and Davison E. Soper${}^{c}$
  }

\address{  \null }

\address{
${}^{a}$Department of Physics,
 University of Kansas, Lawrence, KS  66045   }
\address{${}^{b}$Department of Physics,
 Southern Methodist University,
   Dallas, Texas  75275 }
\address{
  {${}^{c}$}Institute of Theoretical Science,
University of Oregon, Eugene, OR  97403 }


\maketitle

\begin{abstract}
Measurement of the distribution of hadronic energy in the final state in
deeply inelastic electron scattering at HERA can provide a good test of
our understanding of perturbative QCD.   For this purpose, we consider
the energy distribution function, which can be computed without needing
final state parton fragmentation functions. We compute this
distribution function for finite transverse momentum ${q_T}$ at order
$\alpha_s$, and use the results to sum  the perturbation series to
obtain a result valid  for both large and small values of transverse
momentum.

\end{abstract}
\narrowtext

\section{Introduction}

This paper concerns the energy distribution in the final state of
deeply inelastic lepton scattering.  Using a naive parton model, one
would predict that the  scattered parton  appears as a single narrow
jet at a certain angle $(\theta_*,\phi_*)$ in the detector.  Taking
hard QCD interactions into account, one predicts a much richer
structure for the final state energy distribution.  In a previous
paper\citex{mos} (henceforth referred to as~I), we investigated this
structure, using an energy distribution function defined in analogy to the
energy-energy correlation function in $e^+ e^-$
annihilation.\citex{energycor,ellis}
 We studied
this energy distribution as a function of angle $(\theta_B,\phi_B)$
in the detector in the region not too near to the direction
$(\theta_*,\phi_*)$.  In this region, simple QCD perturbation theory is
applicable, and we presented calculations at order $\alpha_s$.  In this
paper, we extend the analysis to the region of $(\theta_B,\phi_B)$
near to $(\theta_*,\phi_*)$ .  Here, multiple soft gluon radiation is
important.  Thus we use a summation of perturbation theory.

\subsection{The energy distribution function}

There is extensive literature on semi-inclusive deeply inelastic
scattering;\citex{early,late,todd,mirkes,graudenz,peccei,morfin}
a brief history and complete set of references can be found
in paper~I.  We begin here with a concise review of how the energy
distribution function is defined, and then discuss how we sum the
contributions that are  important in the  region
$(\theta_B,\phi_B) \simeq (\theta_*,\phi_*)$ to obtain a result which
is valid for all values of $(\theta_B,\phi_B)$.

The reaction that we study is $e+ {\rm A}  \to e +{\rm B} + {\rm X}$ at
the HERA electron-proton collider.\citex{wsmith}
 Let us describe the particles by
their energies and angles in the HERA laboratory frame, with the
positive z-axis chosen in the direction of the proton beam and the
negative z-axis in the direction of the electron beam.  In completely
inclusive deeply inelastic scattering, one measures only $E'$ and
$\theta'$, the energy and angle of the scattered electron.  In the
semi-inclusive case studied in this paper, one also measures some basic
features of the hadronic final state. In principle, one can measure the
energy $E_B$ and the angles $(\theta_B,\phi_B)$ of the outgoing hadron
B.  However, it is much simpler to perform a purely calorimetric
measurement, in which only the total energy coming into a calorimeter
cell at angles $(\theta_B,\phi_B)$ is measured.  This calorimetric
measurement gives the energy distribution
\begin{equation}
{d \Sigma \over
dE'\ d\cos\theta'\ d\cos\theta_B\ d\phi_B}
=
\sum_B \int dE_B\ (1 - \cos\theta_B) E_B\
{d \sigma (e + A \to B + X)
 \over
dE'\ d\cos\theta'\ dE_B\ d\cos\theta_B\ d\phi_B}  \ .
\label{eq:edist}
\end{equation}
The sum runs over all species of produced hadrons B.  We have included
a factor $(1 - \cos\theta_B)$ in the definition because this factor is
part of the  Lorentz invariant dot product $P_{A,\mu} P_B^\mu = E_A E_B
(1 - \cos\theta_B)$.

Notice that $d\Sigma$ measures the distribution of energy in the final
state as a function of angle  without asking how that energy is split
into individual hadrons moving in the same direction.\citex{feynman}
For this reason,
the theoretical expression for $d\Sigma$ will not involve parton decay
functions that describe how partons decay into hadrons.

\subsection{Partonic variables and their relation to HERA lab frame}

At the Born level, the hard scattering process for the reaction  is
$electron + quark \to electron +  quark$ by means of virtual photon or
${\rm Z}_0$ exchange. At order $\alpha_s$, one can have virtual
corrections to the Born graph.  In addition, one can have processes in
which there are two scattered partons in the final state. Then the
initial parton can be either a quark (or antiquark) or a gluon,  while
the observed hadron can come from the decay of either of the final
state partons. Some of these possibilities are illustrated in \xfig{diai},
\xfig{diaii}, and \xfig{diaiii}.

\figdiagi  
\figdiagii  
\figdiagiii  

Let us consider the effect of the emission of the additional,
unobserved, ``bremsstrahlung'' parton.  We can define the part of the
vector boson momentum $q^\mu$ that is transverse to the momentum of the
incoming hadron $P_A^\mu$ and to the momentum of the outgoing hadron
$P_B^\mu$.  One merely subtracts from $q^\mu$ its projections along
$P_A^\mu$ and $P_B^\mu$ (taking  $P_A^2 = P_B^2 = 0$),
\begin{eqnarray}
q_T^\mu &=&
q^\mu
- \dfrac{q \cdot P_B}{P_A \cdot P_B} \, P_A^\mu
- \dfrac{q \cdot P_A}{P_A \cdot P_B} \, P_B^\mu
\quad .
\end{eqnarray}
We let ${q_T} = [- q_T^\mu \cdot {{q_T}}_{\mu}]^{1/2}$ represent the
magnitude of the  transverse momentum. It is ${q_T}$ that is analogous to
the  transverse momentum of produced W's and Z's or lepton pairs in the
Drell Yan process. In the naive parton model, there are no
bremsstrahlung partons and all parton momenta are exactly collinear
with the corresponding hadron momenta, so one has ${q_T} = 0$.  At
order $\alpha_s$, unobserved parton emission allows ${q_T}$ to be
nonzero.

In order to properly describe the parton kinematics we need four more
variables besides ${q_T}$. Two are the standard variables for deeply
inelastic scattering, $Q^2 = -q^\mu q_\mu$ and $x  = Q^2/(2q\cdot
P_A)$. The third is a momentum fraction for the outgoing hadron $B$,
\begin{equation}
z = { P_B\cdot P_A \over q\cdot P_A}  \qquad.
\end{equation}
(Thus the integration over the energy of hadron $B$ in
definition~(\ref{eq:edist}) of the energy distribution is equivalent to an
integration over $z$.) The fourth variable is an azimuthal angle
$\phi$. To define $\phi$, we choose a frame, called the hadron frame,
\xfig{hadframe}, in which the incoming hadron $A$ has its three-momentum
$P_A$ along the positive $z$-axis and the virtual photon four-momentum
$q^{\mu}$
lies along the negative $z$-axis. Then as long as ${q_T}\not=0$, hadron $B$
has some transverse momentum, and we align the $x$- and $y$-axes so that
$P_B^x>0$ and $P_B^y = 0$. We now
define $\phi$ as the azimuthal angle of the incoming lepton in the
hadron frame. These variables are described more fully in paper~I, and
relevant formulas are given in the Appendix of this paper.

\fighadronframe 
\figkini  
\figqtphi       

The variables ${q_T}$ and $\phi$ can be translated to the observables
of the HERA lab frame, \xfig{figkin1}.  In the naive parton model, the
outgoing hadron
$B$ (along with all the other hadrons arising from the  decay of the
struck quark) emerges in the plane defined by the incoming and  outgoing
electrons at a precisely defined angle  $(\theta_*,\phi_*)$, which
can be  computed from the incoming particle momenta and the momentum of
the scattered  electron. The point ${q_T} = 0$ corresponds to
$(\theta_B,\phi_B) = (\theta_*,\phi_*)$.
 We choose our $x$-axis such that $\phi_*=0$.
 Lines of constant positive
${q_T}$ are curves in the $(\theta_B,\phi_B)$ plane that encircle the
point $(\theta_*,\phi_*)$. Lines of constant $\phi$ radiate out of the
point $(\theta_*,\phi_*)$, crossing the lines of constant ${q_T}$.
This is illustrated in \xfig{phietaii}. The precise formulas for the map
relating
$(\theta_B,\phi_B)$ and $({q_T},\phi)$ are given in the Appendix.

In paper~I and in this paper, we find it convenient to convert from the
laboratory  frame variables $\{E^{\prime}, \theta^{\prime}\}$ of the
scattered lepton and $\{\theta_B,\phi_B\}$ of the observed hadron to
$\{x,Q^2\}$ for the lepton and $\{{q_T},\phi\}$ for the observed
hadron. We also convert from $E_B$ to $z$. With this change of
variables, \eq{edist} becomes
\begin{equation}
{ d \Sigma \over dx \, dQ^2 \, d{q_T^2}\, d\phi}
=
\sum_B\int \, dz \, z \ \left(\dfrac{Q^2}{2 x \, E_A}\right) \
{ d \sigma \over dx \, dQ^2 \, d{q_T^2}\, d\phi \, dz}
\ .
\end{equation}
%

\subsection{The Sudakov summation of logarithms of ${q_T}$ }

The main object of study in this paper is the distribution of energy
as a function of ${q_T}$ for ${q_T^2} \ll Q^2$. In paper~I, we applied
straightforward perturbation theory to analyze the energy
distribution in the region ${q_T^2} \sim Q^2$, and  $\alpha_s(Q^2) \ll 1$.
Here there is a rich structure as a function of the angles that relate
the hadron momenta to the lepton momenta.  In fact, a complete
description requires nine structure functions.

When one examines the region ${q_T^2} \ll Q^2$, one finds that the
angular structure simplifies greatly. However, the dependence on ${q_T}$
becomes richer than the dependence on ${q_T}$ of the lowest order graphs.
By summing the most important parts of graphs at arbitrarily high
order, one finds a structure that is sensitive to the fact that QCD is
a gauge theory.

Briefly, the physical picture\citex{parisi} is as follows. At the
Born level of deeply inelastic scattering, a quark in the incoming
proton enters the scattering with momentum $\xi P_A^\mu$ that is
precisely along the beam axis. This quark is scattered by a virtual
photon,
$Z$- or
$W$-boson. Its momentum $\xi P_A^\mu + q^\mu$ is in a direction
$(\theta_*,\phi_*)$ that can be reconstructed by knowing the lepton
momenta. However at higher orders of perturbation theory, the
momentum of the final state parton is
\begin{equation}
(\xi P_A^\mu + q^\mu) - (k_1^\mu + k_2^\mu +\cdots + k_N^\mu)
\qquad ,
\end{equation}
where the $k_i^\mu$ are momenta of gluons that emitted in the
process. In a renormalizable field theory, it is very easy to emit
gluons that are nearly collinear to either the initial or final parton
directions. In addition, in a gauge theory such as QCD, it is very
easy to emit gluons that are soft ($k^\mu \ll Q$). Each gluon
emission displaces ${q_T}$ by a small amount, so that one may think of
the parton as undergoing a random walk in the space of transverse
momenta. With one gluon emission, one finds a cross section that is
singular as ${q_T} \to 0$:
\begin{equation}
{d\sigma \over d\, q_T^2}\
\propto \
\alpha_s\,{a + b \log(q_T^2/Q^2) \over q_T^2}\ .
\end{equation}
At order $\alpha_s^N$ the $1/q_T^2$ singularity is multiplied by a
polynomial in $\log(q_T^2/Q^2)$ of order $2N-1$. This series sums to
a function of $q_T$ that is peaked at $q_T = 0$ but is not singular
there. The width of this distribution is much bigger than the $300\
{\rm MeV}$ that one would guess based on experience with soft hadronic
physics. On the other hand the width is quite small compared to the
hard momentum scale $Q$.

Essentially this same physics has been studied in the two crossed
versions of the process $e + A \to e + B + X$ that can be studied at
HERA.  In electron-positron annihilation, $e + \bar e \to A + B  +
X$, one looks at the energy-energy correlation function for hadrons
$A$ and $B$ nearly back-to-back.\citex{CSS,eedata,LEPSLC}
 In $A+B\to \ell + \bar{\ell} + X$, one studies the distribution of
the lepton pair as a function of its transverse momentum $q_T$
with respect to the beam
axis.\citex{CSS,DWS,drellyan,qiu,fnalexp,naexp,MirkesOhnemus}
 The same analysis applies also to the distribution of the transverse
momentum of $W$ or $Z$ bosons produced in hadron
colliders.\citex{altarelliwzpt,wzpt,ArnoldKauffman,yuanladinsky}

{}From these studies, the following picture emerges. First,
the leading logs ($n=2N-1$) can be summed to all orders,
and dominate the perturbation theory in the  region  $\alpha_s(Q^2) \ll
1$  and  $\alpha_s(Q^2) \, \ln^2(Q^2/{q_T^2})
{\lower0.5ex\hbox{$\stackrel{<}{\sim}$}} 1$. Unfortunately,
most of the interesting physics, and most of the data, lie outside this
region of validity of the leading logarithm approximation. Fortunately,
one can go beyond the leading logarithm summation to obtain a result
that is  valid even when $\alpha_s(Q^2)
\ln^2(Q^2/{q_T^2})$ is large.\citex{CSS,ArnoldKauffman}

The plan for the remainder of the paper is as follows.
 In Sec.~2, we
use our $\alpha_s$ calculation in paper~I to calculate the
asymptotic form of the energy distribution functions in the
${q_T}\to 0$ limit.
 In Sec.~3, we
introduce the Sudakov form factor which sums the
soft gluon radiation in the limit ${q_T}\to 0$.
 In Sec.~4, we
compare the asymptotic  form of the energy distribution functions to
extract the order $\alpha_s$ contributions to the perturbative
coefficients $A$, $B$,
$C^{\rm IN}$ and $C^{\rm OUT}$.
 In Sec.~5, we
address the issue of matching the small ${q_T}$ region to the large ${q_T}$
region.
 In Sec.~6 we investigate the form of the non-perturbative corrections
in the small ${q_T}$ region, and relate these to the Drell-Yan and $e^+e^-$
processes.
 In Sec.~7, we review the principle steps in the calculation.
 In Sec.~8, we
present results for the energy distribution functions throughout the
full ${q_T}$ range.
 Conclusions are presented in Sec.~9, and the Appendix contains a
 set of  relevant formulas.

\section{The  Energy Distribution Functions \label{ECF}}

In this section we review the order $\alpha_s$ perturbative results of
paper~I in order to extract the terms in
$d\Sigma/ dx\, dQ^2\, dq_T^2\, d\phi$ that behave like $1/q_T^2$ times
logs as $q_T \to 0$. In Sec.~3, we display the structure of $d\Sigma/
dx\, dQ^2\, dq_T^2\, d\phi$ with the Sudakov summation of logarithms.
Then in  Sec.~4, by comparing the summed form with the order $\alpha_s$
form of $d\Sigma/ dx\, dQ^2\, dq_T^2\, d\phi$, we will be able to extract
the coefficients that appear in the summed form.

\subsection{Energy Distribution Formulas}

 The process we consider is
$e^- + A \to e^- + B + X$, and
the fundamental formula for the energy distribution is:
\begin{eqnarray}
{ d \Sigma \over dx \, dQ^2 \, d{q_T^2}\, d\phi}
&=&
\sum_{k=1}^{9}  \
{\cal A}_k(\psi,\phi)
\sum_{V_1,V_2}  \
\sum_{j, j'} \
\Sigma_0(Q^2;V_1,V_2,j, j',k) \
\Gamma_k(x,Q^2,q_T^2;j,j')
\qquad .
\label{eq:master}
\end{eqnarray}
 The hyperbolic boost angle, $\psi$, that connects the natural hadron
and lepton frame is given by\citex{angulartheory}
\begin{eqnarray}
{{\rm cosh}} \psi
&=&
\frac{2 x s}{Q^2} - 1
\qquad ,
\end{eqnarray}
and $\phi$ is the azimuthal angle in the hadron  frame.
 The nine angular functions  ${\cal A}_k(\psi,\phi)$
arise from  hyperbolic $D^1(\psi,\phi)$ rotation matrices.
The complete set of ${\cal A}_k(\psi,\phi)$ are listed in the Appendix,
but the two we shall focus on are
\begin{eqnarray}
{\cal A}_1(\psi,\phi) &=& 1 + {{\rm cosh}}^2(\psi) \nonumber\\
{\cal A}_6(\psi,\phi) &=& 2 \  {{\rm cosh}}(\psi)
 \qquad .
\end{eqnarray}
We sum over the intermediate vector bosons $\{V_1, V_2\}=
\{\gamma,Z^0\}$ or $\{W^\pm\}$, as appropriate, and we also sum over
the initial and final partons, $\{j, j'\}$.
 The factor $\Sigma_0(Q^2;V_1,V_2,j, j',k)$ contains the leptonic and
partonic  couplings, the boson propagators, and numerical factors; it is
defined  in the Appendix, \eq{sigzero}.
 It is the hadronic energy distribution  functions,
$\Gamma_k(x,Q^2,q_T^2;j,j')$,
that we shall calculate.

 If we expand the $\Gamma_k$ in the form of
perturbative coefficients convoluted with parton distribution
functions, then  two of the functions $\Gamma_k$, namely $\Gamma_1$
and $\Gamma_6$,  behave like $\log^n({q_T^2}/Q^2)/{q_T^2}$
with $n \ge 0$ for ${q_T}\to 0$. The others behave like $1/{q_T}$ or $1$
times possible logarithms. In this paper we are interested in small
${q_T}$ behavior, so we concentrate our attention on $\Gamma_1$ and
$\Gamma_6$.

What of the less singular structure functions $\Gamma_2$,
$\Gamma_3$, $\Gamma_4$, $\Gamma_5$, $\Gamma_7$, $\Gamma_8$ and
$\Gamma_9$? Fixed order perturbation theory is not applicable for
the calculation of these $\Gamma_k$ for small $q_T$. We note, on the
grounds of analyticity, that these $\Gamma_k$ must be finite or, for
certain $k$, vanish as $q_T \to 0$, even though they have weak
singularities in finite order perturbation theory.  Our perturbative
results in the region of moderate $q_T$ indicate that the fraction
of $d \Sigma/dx\,dQ^2\,dq_T^2\,d\phi$ contributed by these
$\Gamma_k$ is small and dropping as $q_T$ decreases. We thus
conclude that these contributions would be hard to detect
experimentally for small $q_T$. For this reason, we do not address
the problem of summing perturbation theory for  $\Gamma_2$,
$\Gamma_3$, $\Gamma_4$, $\Gamma_5$, $\Gamma_7$, $\Gamma_8$ and
$\Gamma_9$.

Applying the methods of Refs.~\cite{CS,CSS} to deeply inelastic
scattering, we write $\Gamma_1$ in the form
\begin{eqnarray}
\Gamma_1(x,Q^2,{q_T^2};j,j') &=&
 \Gamma_1^{Pert}(x,Q^2,{q_T^2};j,j')
-  \Gamma_1^{Asym}(x,Q^2,{q_T^2};j,j')
+ W(x,Q^2,q_T^2;j,j')
\quad .
\nonumber \\
\label{eq:smallqTdecompose}
\end{eqnarray}
Here $W(x,Q^2,q_T^2;j,j')$ sums the singular terms to all orders, and
contains the leading behavior of $\Gamma_1$ as ${q_T} \to 0$.
 $\Gamma_1^{Pert}(x,Q^2,{q_T^2};j,j')$ is simply $\Gamma_1(x,Q^2,{q_T^2};j,j')$
evaluated at a finite order ($\alpha_s^1$ for our purpose) in perturbation
theory.
 $\Gamma_1^{Asym}(x,Q^2,{q_T^2};j,j')$ equals $W(x,Q^2,q_T^2;j,j')$ truncated
at a finite order of $\alpha_s$ in perturbation theory.
 Specifically, if we expand $W(x,Q^2,q_T^2;j,j')$ in the form
of perturbative coefficients convoluted with parton distribution functions,
then the coefficients have the form of $\log^n({q_T^2}/Q^2)/{q_T^2}$ with $n
\ge
0$. There are, by definition, no terms that behave like
$({q_T^2}/Q^2)^p$ times possible logarithms for $p > -1$. Such terms
exist in $\Gamma_1$, but they are associated with
$(\Gamma_1^{Pert} -  \Gamma_1^{Asym})$  in \eq{smallqTdecompose}.

The angular function ${\cal A}_1(\psi,\phi)= 1 + {{\rm cosh}}^2(\psi)$
that multiplies $W$ in
the small ${q_T}$ limit arises from the numerator factor
\begin{equation}
{\rm Tr} \{ {{\ell \kern -4.6 pt \raise 2 pt \hbox{/}}} \gamma_\mu {{\ell \kern
-4.6 pt \raise 2 pt \hbox{/}}}^{\prime} \gamma_\nu\}\
{\rm Tr} \{ {{P \kern -6.5 pt \raise 2 pt \hbox{/}}}_{\!\!A} \gamma^\mu {{P
\kern -6.5 pt \raise 2 pt \hbox{/}}}_{\!\! B} \gamma^\nu\}
\qquad .
\end{equation}
Here
${\rm Tr} \{{{\ell \kern -4.6 pt \raise 2 pt \hbox{/}}}
\gamma_\mu {{\ell \kern -4.6 pt \raise 2
pt \hbox{/}}}^{\prime} \gamma_\nu\}$ is associated with the lepton scattering,
and the factor
${{P \kern -6.5 pt \raise 2 pt \hbox{/}}}_{\!\!A}
\cdots {{P \kern -6.5 pt \raise 2 pt \hbox{/}}}_{\!\! B}$
gives the Dirac structure of
the hadronic part of the cut diagram in
\xfig{nonpertiii}(b) in the limit ${q_T} \to 0$.
 We will discuss \xfig{nonpertiii}
further in Sec.~\ref{NONPERT}.

\fignonpertiii       

The weak currents also contain $\gamma_5\gamma^\mu$ terms. This gives
the possibility of another angular function in the small ${q_T}$
limit. With the same limiting hadronic structure,
${{P \kern -6.5 pt \raise 2 pt \hbox{/}}}_{\!\!A}\cdots {{P \kern -6.5 pt
\raise 2 pt \hbox{/}}}_{\!\! B}$ we can have
\begin{equation}
{\rm Tr} \{ {{\ell \kern -4.6 pt \raise 2 pt \hbox{/}}} \gamma_5\gamma_\mu
{{\ell \kern -4.6 pt \raise 2 pt \hbox{/}}}^{\prime}
\gamma_\nu\}\
{\rm Tr} \{ {{P \kern -6.5 pt \raise 2 pt \hbox{/}}}_{\!\!A} \gamma_5\gamma^\mu
 {{P \kern -6.5 pt \raise 2 pt \hbox{/}}}_{\!\! B}
\gamma^\nu\}
 ,
\nonumber \\
\end{equation}
which is proportional to the angular function
${\cal A}_6(\psi,\phi) = 2 \, {{\rm cosh}}(\psi)$
at ${q_T} = 0$.
 (Note that both
${\cal A}_1(\psi,\phi)$ and ${\cal A}_6(\psi,\phi)$
are independent of the azimuthal angle $\phi$.)
 Thus $\Gamma_6$ has the structure
\begin{eqnarray}
\Gamma_6(x,Q^2,{q_T^2};j,j') &=&
 \Gamma_6^{Pert}(x,Q^2,{q_T^2};j,j')
-  \Gamma_6^{Asym}(x,Q^2,{q_T^2};j,j')
+ (-1) W^{}(x,Q^2,q_T^2;j,j')
 ,
\nonumber \\
\label{eq:smallqTdecomposeii}
\end{eqnarray}
with the same function\footnote{
The minus sign in front of $W(x,Q^2,{q_T^2};j,j')$ in
\eq{smallqTdecomposeii} arises from our convention
for the functions
${\cal A}_k(\psi,\phi)$ and  couplings
$\Sigma_0(Q^2;V_1,V_2,j, j',k)$ that multiply
$\Gamma_1$
and $\Gamma_6$.
}
 $W$ as in  \eq{smallqTdecompose}.
 Again
$W$ contains the terms that behave like $\log^n({q_T^2}/Q^2)/{q_T^2}$ in
perturbation theory, while $(\Gamma_6^{Pert} -  \Gamma_6^{Asym})$ contain
the less singular terms. Our object now will be to study the small ${q_T}$
function $W$.

\subsection{Parton Level Distributions}

The above  hadronic process takes place via the partonic sub-process
$V(q) + a(k_a) \to b(k_b) + X $
where $V$ is an intermediate vector boson, and
$a$ and $b$ denote parton species.
 The  hadron structure function $W(x,Q^2,{q_T^2};j,j')$ is
related to a perturbatively
calculable  parton level structure function
$w_a({\widehat x },Q^2,{q_T^2};j,j')$ via
\begin{eqnarray}
W(x,Q^2,{q_T^2};j,j') &=&
f_{a/A} \otimes  w_a
=
\int_x^1 { d\xi \over \xi} \
\sum_a f_{a/A}(\xi,\mu) \
w_a({\widehat x },Q^2,{q_T^2};j,j')
\qquad ,
\label{eq:partonhadron}
\end{eqnarray}
with $\xi_a = k_a^+/P_A^+$ and ${\widehat x }=x/\xi_a$.
 Here $f_{a/A}$ is the
{\vbox{\hrule\kern 1pt\hbox{\rm MS}}}\ parton distribution function.
 Note that the decay distribution function $d_{B/b}(\xi_b,\mu)$ is
absent since we have used the extra $\int z \, dz$
and the sum over hadrons
from the definition of
the
energy distribution to  integrate out the $d_{B/b}(\xi_b,\mu)$
via the momentum sum rule,
\begin{eqnarray}
\sum_{B} \  \int \ d\xi_b \ \xi_b \ d_{B/b} (\xi_b,\mu) &=& 1
\qquad .
\label{}
\end{eqnarray}

 The  partonic  structure
function, $w_a({\widehat x },Q^2,{q_T^2};j,j')$,
is obtained by first computing the partonic tensor
\begin{eqnarray}
w^{\mu\nu}(k_a,k_b,q) &=&
\dfrac{1}{2} \,
\sum_{X,s,s^\prime} \
\int \
d^4 x \ e^{-i q\cdot x} \
\langle  k_a,s | j^\nu(0) | k_b, s^\prime; X   \rangle \
\langle  k_b, s^\prime; X |  j^\mu(0) |  k_a,s \rangle
 ,
\end{eqnarray}
which is a matrix element of current operators.
We then project out the appropriate angular component
({\it cf.}, paper~I), and extract the leading term in the
${q_T}\to 0$ limit.
 Explicit calculation will show that these limits
(up to overall factors) are
identical for the projection of the 1 and 6 tensors.
 In the small ${q_T}$ limit, the energy distribution function
is then given by:
\begin{eqnarray}
{d \Sigma \over
dx\ dQ^2\ d{q_T^2}\ d\phi}
&\simeq&
{\cal A}_1(\psi,\phi) \
\sum_{V_1,V_2} \
\sum_{j, j'} \
\Sigma_0(Q^2;V_1,V_2,j, j',1) \
\sum_a \
f_{a/A}(\xi,\mu) \otimes
w_a({\widehat x },Q^2,{q_T^2};j,j')
 \nonumber \\ &-&
{\cal A}_6(\psi,\phi) \
\sum_{V_1,V_2} \
\sum_{j, j'} \
\Sigma_0(Q^2;V_1,V_2,j, j',6) \
\sum_a \
f_{a/A}(\xi,\mu) \otimes
w_a({\widehat x },Q^2,{q_T^2};j,j')
 \nonumber \\ &+&
{\rm \ plus\ terms\ less\ singular\ than\   1/{q_T^2}}
\qquad .
\end{eqnarray}
Again, the relative minus sign is simply due to the definition of
${\cal A}_k(\psi,\phi)$ and $\Sigma_0(Q^2;V_1,V_2,j, j',k)$.

\subsection{The Asymptotic Energy Distribution Functions}

We observe (from the
results  of paper~I) that the perturbative
$\Gamma_1^{Pert}(x,Q^2,q_T^2;j,j')$ and
$\Gamma_6^{Pert}(x,Q^2,q_T^2;j,j')$  diverge as $1/{q_T^2}$ for ${q_T} \to 0$.
 To identify the singular terms, we can
expand the  on-shell delta function for small ${q_T}$ using
\begin{eqnarray}
\ 2\pi\ \delta\!\left[ (q^\mu + k_a^\mu - k_b^\mu)^2\right]
&=&
{ 2 \pi {\widehat x } \over Q^2} \
\Biggl\{
\ln \left( {Q^2 \over {q_T^2}} \right)  \delta(1-{\widehat x }) \
\delta(1-{\widehat z })
+
{  \delta(1-{\widehat z }) \over (1-{\widehat x })_+ }
+
{  \delta(1-{\widehat x }) \over (1-{\widehat z })_+ }
\Biggr\}
 ,
\nonumber \\
\label{}
\end{eqnarray}
where the ``$+$"-prescriptions is defined as usual by:
\begin{eqnarray}
\int_z^1 dy \ \dfrac{G(y)}{(1-y)_+}
&=&
G(1) \ln(1-z) +
\int_z^1 dy \ \dfrac{[G(y)-G(1)]}{(1-y)}
\qquad .
\label{}
\end{eqnarray}
  Taking the ${q_T} \to 0$ limit for the results of paper~I,
we find the partonic energy distribution to be
\begin{eqnarray}
w_a^{Asym}({\widehat x },Q^2,{q_T^2};j,j')
&=&
\left[
\dfrac{16 \pi^2 \, \alpha_s}{{q_T^2}}
\right] \
\Biggl\{
\delta_{a,j} \, \delta(1-{\widehat x }) \ C_F \
\left[ 2 \ln \left(\dfrac{Q^2}{{q_T^2}} \right) - 3 \right]
\nonumber \\[10pt]
&+&   \delta_{a,j} \ C_F \ \left[ \dfrac{1+{\widehat x }^2}{1-{\widehat x }}
\right]_+ \
 +    \delta_{a,g} \ \left[ \dfrac{{\widehat x }^2+ (1-{\widehat x })^2}{2}
\right]
\Biggl\}
\qquad ,
\label{eq:eqbmu}
\end{eqnarray}
where we use  $\delta_{a,j}$ and $\delta_{a,g}$ for the quark and gluon
contributions, respectively.
 For convenience, we denote the asymptotic limit ${q_T}\to 0$ of
$w_a$  by $w_a^{Asym}$.

  In this limit, we can greatly simplify this expression by
identifying the QCD splitting functions. We present the result for the
hadronic structure function convoluted with the parton distributions,
({\it cf.}, \eq{partonhadron}):
\begin{eqnarray}
\Gamma^{Asym}(x,Q^2,{q_T^2};j,j')
&=&
\left[
\dfrac{16 \pi^2 \, \alpha_s}{{q_T^2}}
\right] \
\Bigl\{
f_{j/A}(x) \ C_F \
\left[
2  \ \ln \left(\dfrac{Q^2}{{q_T^2}} \right) - 3
\right]
\nonumber  \\
&+&
  f_{j/A} \otimes P_{q/q}
+ f_{g/A} \otimes P_{q/g}
\Bigr\}
\qquad ,
\label{eq:sudexp}
\end{eqnarray}
where $\otimes$ represents a convolution in ${\widehat x }$.
 In the simple form above, it is easy to identify the separate
contributions.
 The last two terms arise from the collinear singularities, and are
proportional to
the appropriate first order splitting kernel, $P_{q/q}$ and $P_{q/g}$.
 It is the remaining term in which we are interested as these arise from
the
soft gluon processes.
 We note that
$\Gamma^{Asym}$ is defined such that the combination
$\Gamma^{Pert}-\Gamma^{Asym}$
has only logarithmic singularities
 as ${q_T}\to 0$.

\section{Sudakov Form Factor}

In this section, we display the structure of $d\Sigma/ dx\, dQ^2\,
dq_T^2\, d\phi$ with the Sudakov summation of logarithms. This provides
the basis a formula that includes nonperturbative effects, developed in
Sec.~6. In addition, in Sec.~4 we compare the summed form of this
section with the order $\alpha_s$ form of $d\Sigma/ dx\, dQ^2\,
dq_T^2\, d\phi$ from Sec.~2, in order to extract the coefficients that
appear in the summed form.

\subsection{Bessel Transform
of $w_a({\widehat x },Q^2,{q_T^2};j,j')$}

It proves
convenient to introduce a Fourier transform  between transverse momentum
space (${q_T}$) and impact parameter space ($b$),
 \begin{eqnarray}
w_a({\widehat x },Q^2,{q_T^2};j,j') &=&
\int {d^2 b \over (2\pi)^2}
\  e^{i {q_T} \cdot b }  \
\widetilde{w}_a({\widehat x },Q^2,b^2;j,j')
\nonumber \\
&=&
\int_0^{\infty} {d b \over 2\pi} \ b \ J_0(b \, {q_T}) \
\widetilde{w}_a({\widehat x },Q^2,b^2;j,j')
\qquad ,
\label{eq:btransform}
\end{eqnarray}
 as
$\widetilde{w}_a({\widehat x },Q^2,b^2;j,j')$ will have a simple
structure.\citex{CSS}
 Effectively, we make use of the renormalization group equation
to sum the logs of $Q^2$, and gauge invariance to sum the logs of ${q_T}
\sim 1/b$.
 The Fourier transform also maps the ${q_T}$ singularities at the
origin to the large $b$ behavior of
$\widetilde{w}_a({\widehat x },Q^2,b^2;j,j')$; we will take advantage of
this when we consider non-perturbative contributions.

\subsection{Sudakov Form Factor}

The structure function in impact parameter space
 $\widetilde{w}_a({\widehat x },Q^2,b^2;j,j')$ has the factorized form:
\begin{eqnarray}
\widetilde{w}_a({\widehat x },Q^2,b^2;j,j')
&=&
C^{IN }_{ja} \big({\widehat x } ,b\mu \big) \
\sum_{a'}
\int \, d{\widehat z } \,  {\widehat z } \
C^{OUT}_{a^{\prime}\, j^{\prime}} \big({\widehat z } ,b\mu \big) \
e^{-S(b)}
\qquad .
\label{eq:suddefi}
\end{eqnarray}
This form is from references \cite{CS} and \cite{CSS} applied to the DIS
process, and generalized to include vector bosons other than the photon.
 The last
exponential factor is the Sudakov form factor:
\begin{eqnarray}
S(b)
&=&
\int_{C_1^2/b^2}^{C_2^2 Q^2} \
{d\mu^2 \over \mu^2}
\left\{\ln\left[{C_2^2 Q^2\over \mu^2}\right]
A( \alpha_s(\mu) ) +
B( \alpha_s(\mu) )\right\}
\qquad .
\label{eq:suddefii}
\end{eqnarray}
 The logarithm in the exponential is characteristic of the
gauge theory. It arises from the soft gluon summation in QCD at the low
transverse  momentum ${q_T^2} \ll Q^2$.
 The arbitrary constants $\{ C_1, C_2 \}$ reflect the freedom in the choice of
renormalization scale.
 We  choose  $\{ C_1, C_2 \}$  to be
\begin{eqnarray}
C_1 &=& 2 e^{-\gamma_E}   \\
C_2 &=& 1   \qquad .
\label{eq:c1c2}
\end{eqnarray}
 The functions $A$, $B$ and the
hard scattering  functions
$C$'s are simple power series in the  strong coupling constant
$\alpha_s$ with  numerical
coefficients:\footnote{
Collins and Soper\citex{CSS} (CS)
expand in powers of ${\alpha_s}/\pi$, and
 Davies, Webber, and Stirling\citex{DWS} (DWS)
expand in powers of ${\alpha_s}/(2\pi)$.
 We carry the extra factor of (2)
explicitly  to facilitate comparison between these references.}
\begin{eqnarray}
A( \alpha_s(\mu) ) &=&
\sum_{N=1}^\infty\   \left\{{\alpha_s(\mu)\over (2) \pi}\right\}^N \ A_N
\\
B( \alpha_s(\mu) ) &=&
\sum_{N=1}^\infty\   \left\{{\alpha_s(\mu)\over (2) \pi}\right\}^N \ B_N
\label{eq:abterms}
\end{eqnarray}
\begin{eqnarray}
C_{ja}^{IN}( {\widehat x },b\mu ) &=&
\delta(1-{\widehat x }) \, \delta_{ja} + \sum_{N=1}^\infty \
C_{ja}^{IN(N)} ({\widehat x },b\mu ) \
\left\{{\alpha_s(\mu)\over (2) \pi}\right\}^N
\label{eq:ctermsi}
\\
C_{a^{\prime}j^{\prime}}^{OUT}( {\widehat z },b\mu ) &=&
\delta(1-{\widehat z }) \, \delta_{a^{\prime}j^{\prime}} + \sum_{N=1}^\infty \
C_{a^{\prime}j^{\prime}}^{OUT(N)}({\widehat z },b\mu ) \
\left\{{\alpha_s(\mu)\over (2) \pi}\right\}^N
\qquad .
\label{eq:ctermsii}
\end{eqnarray}
The normalization has been chosen such that each hard scattering function
$C$ equals a
\hbox{$\delta$-function} at leading order.

As noted in reference \cite{CSS}, in the limit $Q\to \infty$, all
logarithms may be counted as being equally large. Therefore, to evaluate
the cross section at ${q_T}\simeq0$ to an approximation of ``degree $N$,"
one must evaluate
$A$ to order ${\alpha_s}^{N+2}$,
$B$ to order ${\alpha_s}^{N+1}$,
$C^{IN}$ and $C^{OUT}$ to order ${\alpha_s}^{N}$,
and the $\beta$ function order ${\alpha_s}^{N+2}$.
In particular, an extra order in $A$ is necessary due to the extra
logarithmic factor in \eq{suddefii}.
 For the present calculation, we evaluate
$A$ to order ${\alpha_s}^{2}$,
$B$ to order ${\alpha_s}^{1}$,
$C^{IN}$ and $C^{OUT}$ to order ${\alpha_s}^{1}$,
and the $\beta$ function to order ${\alpha_s}^{2}$.
This yields the cross section
to order ${\alpha_s}^{1}$ for large ${q_T}$,
to order ${\alpha_s}^{0}$ for small ${q_T}$,
and the cross section integrated over ${q_T}$ to
${\alpha_s}^{1}$.

\subsection{Perturbative Expansion of the Sudakov Form Factor}

We can extract the $A_i$ and $B_i$ coefficients of the Sudakov factor by
expanding $\widetilde{w}_a({\widehat x },Q^2,b^2;j,j')$
of \eq{suddefi} in ${\alpha_s}$, and comparing with  the perturbative
calculation of paper~I.
 Here, we take a fixed  momentum scale $\mu_0$ in $\alpha_s(\mu_0)$ as the
running of
$\alpha_s(\mu)$
contributes only to higher orders. We can now compute the integral over
$\mu^2$ analytically to obtain:
\begin{eqnarray}
S(b)
&=&
\int_{C_1^2/b^2}^{C_2^2 Q^2} \
{d\mu^2 \over \mu^2} \
\left[\ln\left[{C_2^2 Q^2\over \mu^2}\right] A( \alpha_s(\mu) ) +
B( \alpha_s(\mu) )\right]
\nonumber\\[5pt]
&\simeq&
{\alpha_s(\mu_0) \over (2) \,  \pi }
\left[
 A_1 \, {L^2 \over 2}
+ B_1 L \right]
\qquad ,
\end{eqnarray}
where
\begin{eqnarray}
L&=& \ln\left[ {C_2^2 \over C_1^2 } \   b^2  \, Q^2\right]
\qquad .
\end{eqnarray}
 We  expand the Sudakov exponential out to order ${\alpha_s}^1$,
\begin{eqnarray}
e^{-S(b)} &\simeq&
1 - S(b) + {\cal O}({\alpha_s}^2)
\qquad ,
\end{eqnarray}
and perform the Bessel transform of
$\widetilde{w}_a({\widehat x },Q^2,b^2;j,j')$
({\it cf.}, \eq{suddefi})
to  obtain the partonic
structure function in momentum space:
 \begin{eqnarray}
w_{a}^{}({\widehat x },Q^2,{q_T^2};j,j')
&=&
\int_0^{\infty} {d b \over 2\pi} \ b \ J_0(b \, {q_T}) \
\left[
\delta_{a,j} \, \delta(1-{\widehat x }) +
  \dfrac{{\alpha_s}(\mu)}{(2) \pi} \  C_{a,j}^{IN\, (1)}({\widehat x },b\mu)
\right]
\nonumber \\ &\times&
\sum_{a'} \
\left[
\delta_{a^\prime,j^\prime}  +
  \int_0^1 \, {\widehat z } \, d{\widehat z } \  \dfrac{{\alpha_s}(\mu)}{(2)
\pi} \
  C_{a^\prime,j^\prime}^{OUT\, (1)}({\widehat z },b\mu)
\right]
\nonumber \\[10pt]  &\times&
\left[
1 - S(b) + {\cal O}({\alpha_s}^2)
\right]
\qquad ,
\end{eqnarray}
where we have used the  first order  expressions for
$C^{IN \, (N)}_{jk} ({\widehat x },\mu b)$
and
$C^{OUT \, (N)}_{jk} ({\widehat z },\mu b)$.

 Finally, we integrate to obtain the ${\cal
O}({\alpha_s}^1)$ terms for finite ${q_T}$:
\begin{eqnarray}
w_a^{}({\widehat x },Q^2,{q_T^2};j,j')
&\simeq&
\left[
\dfrac{16 \pi^2 \, \alpha_s}{{q_T^2}}
\right] \
\Biggl\{
\delta_{a,j}  \,
\delta(1-{\widehat x }) \
\left[
\dfrac{2 A_1}{(2)} \,
\left\{
 \ln \left(\dfrac{Q^2}{{q_T^2}} \right)
-2\ln \left(\dfrac{e^{\gamma_E} C_1}{2 C_2} \right)
\right\}
+
\dfrac{2 B_1}{(2)}
\right]
\,
\nonumber\\[10pt] && \qquad \qquad \qquad
+ \ \delta_{a,j} \ P_{q/q}(x)
+ \ \delta_{a,g} \ P_{q/g}(x)
\nonumber\\ && \qquad \qquad \qquad
+ {\rm \ terms\ proportional\ to\ } \delta({q_T^2})
\Biggl\}
\qquad .
\label{eq:sudexpi}
\end{eqnarray}
Here, we have use the fact that the renormalization group equation tells us
the
form of $C^{IN(1)}({\widehat x },\mu b)$ and $C^{OUT(1)}({\widehat z },\mu b)$
must be
a splitting kernel times $\log[\mu b]$, plus a function independent of $\mu$
and $b$.
Equivalently, for the hadronic structure function, we find:
\begin{eqnarray}
W^{}(x,Q^2,{q_T^2};j,j')
&\simeq&
\left[
\dfrac{16 \pi^2 \, \alpha_s}{{q_T^2}}
\right] \
\Biggl\{
f_{j/A}(x) \
\left[
\dfrac{2 A_1}{(2)} \,
\left\{
 \ln \left(\dfrac{Q^2}{{q_T^2}} \right)
-2\ln \left(\dfrac{e^{\gamma_E} C_1}{2 C_2} \right)
\right\}
+
\dfrac{2 B_1}{(2)}
\right] \,
\nonumber\\[10pt] && \qquad \qquad \qquad
+ \ f_{j/A} \otimes  \ P_{q/q}
+ \ f_{g/A} \otimes  \ P_{q/g}
\nonumber\\ && \qquad \qquad \qquad
+ {\rm \ terms\ proportional\ to\ } \delta({q_T^2})
\Biggl\}
\qquad .
\label{eq:sudexpii}
\end{eqnarray}
 We will compare the first-order expansions in \eq{sudexpi} and
\eq{sudexpii} with the asymptotic limit of the perturbative calculations
of Sec.~\ref{ECF} to extract the desired $A_1$ and $B_1$ coefficients.

\section{Comparing  Asymptotic and Sudakov Contributions}

In this section, we compare the summed form
 of $d\Sigma/ dx\, dQ^2\, dq_T^2\, d\phi$
with the order $\alpha_s$
form, and thus extract the
coefficients that appear in the summed form.

\subsection{Extraction of $A$ and $B$}

Comparing the expansion of the Sudakov expression
[\eq{sudexpii}]
 with the asymptotic  results [\eq{sudexp}], we
obtain the order $\alpha_s^1$ coefficients $A_1$ and $B_1$,
\begin{eqnarray}
A_1 &=& (2) \     C_F  \\
B_1 &=& (2) \ 2\, C_F \,
\ln \left[ {C_1\over 2 C_2} \, e^{\gamma_E -(3/4)} \right]
\qquad .
\end{eqnarray}
 With our particular choice of the arbitrary constants $\{ C_1, C_2 \}$
in \eq{c1c2}, we have:
\begin{eqnarray}
A_1 &=& (2) \     C_F  \\
B_1 &=& (2) \ \left[ \dfrac{-3}{\phantom{-} 2} \right]  \ C_F
\qquad .
\end{eqnarray}
 We find that the results for  $A_1$ and $B_1$ obtained above are
identical to those found  in reference~\cite{CS} for Drell-Yan
production,  as well as those found  in reference~\cite{eedata} for $e^+\,
e^-$ annihilation. This apparent crossing symmetry has been demonstrated at
order
$\alpha_s^2$ by Trentadue.\citex{trentadue} In light of this result,
we shall make use of the $A_2$ coefficient\citex{trentadue}
\begin{eqnarray}
A_2 &=& (4) \left\{
{67\over9} - {\pi^2\over 3}  - {10\over 27} \,  N_f
+ {2\over 9} (33-2N_f) \ln\left({C_1 \over 2 \, e^{-\gamma_E}}   \right)
\right\}
\qquad .
\end{eqnarray}
The extra order in the $A_i$ expansion will compensate extra logarithm
$L$ which is not present for the $B_i$ terms.

\subsection{Expansion of $C^{IN}$ and  $C^{OUT}$}

$C^{IN}$ and  $C^{OUT}$ terms are obtained by comparing the terms in the
perturbative expansion proportional to $\delta({q_T})$
 with the expanded summed form.
 Since the  virtual graphs yield contributions only proportional to
$\delta({q_T})$, they will only enter $C^{IN}$ and  $C^{OUT}$.
 The real graphs  yield {\it both} zero and finite ${q_T}$ terms;
therefore, they will contribute to both $A_i$, $B_i$, and
the  $C^{IN}$ and  $C^{OUT}$ coefficients.
 The calculation of the virtual graphs has been performed by
Meng\rlap,\citex{meng} and we make use of those results.

We have defined the $C ({\widehat x },\mu b)$ coefficients
such that at leading order, they are
\begin{eqnarray}
C^{IN \, (0)}_{jk} ({\widehat x },\mu b)
&=&
\delta_{jk} \, \delta(1-{\widehat x } )
\nonumber \\[10pt]
C^{OUT \, (0)}_{jk} ({\widehat z },\mu b)
&=&
\delta_{jk} \, \delta(1-{\widehat z } )
\nonumber \\[10pt]
C^{IN \, (0)}_{jg} ({\widehat x },\mu b)
&=&
C^{OUT \, (0)}_{gk} ({\widehat z },\mu b)
= 0
\qquad .
\end{eqnarray}
(Here, $j$ and $k$ denote quarks and anti-quarks, and $g$ denotes
gluons.)
 At next to leading order,
we find that $C^{IN \, (1)}  ({\widehat x },\mu b)$
match those calculated by CS for the Drell-Yan process:\citex{CS}
\begin{eqnarray}
C^{IN \, (1)}_{jk} ({\widehat x },\mu b)
&=&
\delta_{jk} \Biggl\{
{2\over 3} (1-{\widehat x })
+P_{q/q}({\widehat x }) \
\ln \left( { \lambda \over   \mu b}     \right)
\nonumber \\[10pt]
&+&  \delta(1-{\widehat x }) \ \Biggl[
-C_F \, \ln^2 \left( {C_1 \ e^{-3/4} \over C_2 \, \lambda } \right)
+ {\pi^2 \over 3} - {23 \over 12}
\Biggr]
\Biggr\}
\end{eqnarray}
\begin{eqnarray}
C^{IN \, (1)}_{jg} ({\widehat x },\mu b)
&=&
{1\over 2} {\widehat x } (1-{\widehat x })
+P_{j/g}({\widehat x }) \
\ln \left( { \lambda \over   \mu b}     \right)
\qquad .
\end{eqnarray}
The $C^{OUT \, (1)}  ({\widehat z },\mu b)$ are
simply those for  $e^+ e^-$ as given in reference~\cite{eedata}:
\begin{eqnarray}
C^{OUT \, (1)}_{jk} ({\widehat z },\mu b)
&=&
\delta_{jk} \Biggl\{
{2\over 3} (1-{\widehat z })
+P_{q/q}({\widehat z }) \
\ln \left( { \lambda \over   \mu b}     \right)
\nonumber \\[10pt]
&+&  \delta(1-{\widehat z }) \ \Biggl[
-C_F \, \ln^2 \left( {C_1 \ e^{-3/4} \over C_2 \, \lambda } \right)
+ {\pi^2 \over 3} - {29 \over 12}
\Biggr]
\Biggr\}
\end{eqnarray}
\begin{eqnarray}
C^{OUT \, (1)}_{gk} ({\widehat z },\mu b)
&=&
{2\over 3} \, {\widehat z }
+P_{g/k}({\widehat z }) \
\ln \left( { \lambda \over   \mu b}     \right)
\qquad ,
\end{eqnarray}
where we define $\lambda=2e^{-\gamma_E}$ to simplify the notation. Note that
$C^{IN}$ and
$C^{OUT}$ are only a function of the ratio $C_1/C_2$.

\subsection{Complete Expression}

Now that we have obtained  $A_1$, $A_2$, and $B_1$, we can substitute into
equation~\eq{suddefii} to obtain the complete Sudakov contribution
(including the full
$\alpha_s(\mu)$
dependence).  We choose to perform the $\mu$ integral analytically, as the
Bessel transform would be prohibitively CPU intensive if we did not.
To facilitate this computation we provide an integral table in
Appendix~\ref{APPA} including all the necessary terms
 We are now ready to combine the separate parts of the calculation.

\section{Matching}

We now have computed the contributions to the energy distribution
functions for the
perturbative $\Gamma_k^{Pert}$ in paper~I [\eq{partonhadron}],
the summed (or Sudakov) $W^{}(x,Q^2,q_T^2;j,j')$ in \eq{sudexpii}, and
the asymptotic $\Gamma_k^{Asym}$ in \eq{sudexp}.
 We can simply assemble these pieces
to form the total structure functions via:
\begin{eqnarray}
\Gamma_1(x,Q^2,{q_T^2};j,j') &=&
 \Gamma_1^{Pert}(x,Q^2,{q_T^2};j,j')
+ \phantom{(-1)} W^{}(x,Q^2,q_T^2;j,j')
-  \Gamma_1^{Asym}(x,Q^2,{q_T^2};j,j')
 \nonumber\\
\Gamma_6(x,Q^2,{q_T^2};j,j') &=&
 \Gamma_6^{Pert}(x,Q^2,{q_T^2};j,j')
+ (-1) W^{}(x,Q^2,q_T^2;j,j')
 -  \Gamma_6^{Asym}(x,Q^2,{q_T^2};j,j')
 .
\nonumber \\
\label{eq:}
\end{eqnarray}
Here, $\Gamma_k^{Pert}$ and $\Gamma_k^{Asym}$ are evaluated
at order ${\alpha_s}^1$, while
$W$ contains a summation of perturbation theory.
 In the limit ${q_T}\to0$,
$\Gamma_k^{Pert}$ and $\Gamma_k^{Asym}$ will cancel
each other leaving  $W$ as we desire.
In the limit ${q_T}\simeq Q$, $W$  and
$\Gamma_k^{Asym}$  will cancel  to leading order in $\alpha_s$;
however, the finite difference may not be
negligible. To ensure that we recover the proper
result ($\Gamma_k^{Pert}$) for large ${q_T}$, we
define the total energy distribution
function ($\Gamma_k$) to be:
\begin{eqnarray}
\Gamma_1(x,Q^2,{q_T^2};j,j') &=&
 \Gamma_1^{Pert}(x,Q^2,{q_T^2};j,j')
\nonumber \\
&+&
 {\cal T}\left(\dfrac{{q_T}}{Q}\right)
\left\{ \phantom{(-1)} W^{}(x,Q^2,q_T^2;j,j') -
\Gamma_1^{Asym}(x,Q^2,{q_T^2};j,j') \right\}
 \nonumber\\
\Gamma_6(x,Q^2,{q_T^2};j,j') &=&
 \Gamma_6^{Pert}(x,Q^2,{q_T^2};j,j')
\nonumber \\
&+&
  {\cal T}\left(\dfrac{{q_T}}{Q}\right)
\left\{(-1) W^{}(x,Q^2,q_T^2;j,j') -  \Gamma_6^{Asym}(x,Q^2,{q_T^2};j,j')
\right\}
\qquad ,
\nonumber \\
\label{eq:}
\end{eqnarray}
where we introduce the arbitrary function
\begin{eqnarray}
{\cal T}\left( {{q_T} \over Q} \right)
&=&
{1 \over 1+
\left(\rho \, \dfrac{{q_T}}{Q}
\right)^4 }
\qquad .
\label{eq:rhoeq}
\end{eqnarray}
The  transition function
${\cal T}({q_T}/Q)$ serves to switch smoothly
from the matched formulas to the perturbative
formula, and
$\rho$ is an arbitrary parameter which determines
the details of the matching.
\xfig{tempvii} displays ${\cal T}({q_T}/Q)$ for a range of $\rho$ values.
We will choose $\rho=5$ which ensures that
$\Gamma_k \simeq \Gamma_k^{Pert}$ for
${q_T}/Q \,  {\lower0.5ex\hbox{$\stackrel{>}{\sim}$}} \, 0.4$, a conservative
value.

\figtempvii 

\section{Non-Perturbative Contributions \label{NONPERT}}

In analogy with \eq{partonhadron} and \eq{suddefi},
the Bessel transform of the hadronic structure function is defined
as:
\begin{eqnarray}
W^{}(x,Q^2,q_T^2;j,j')
&=&
\int \, {d^2 b \over (2\pi)^2 } \
e^{i {q_T} \cdot b } \
\widetilde{W}^{}(x,Q^2,b^2;j,j')
\qquad .
\label{eq:wdefii}
\end{eqnarray}
When $b$ is small, we have:
\begin{eqnarray}
\widetilde W(x,Q^2,b^2;j,j')
&=&
\int_{x}^{1}
\dfrac{d\xi}{\xi} \
\sum_a \
f_{a/A}(\xi,\mu) \
C^{IN }_{ja} \big({\widehat x } ,b\mu \big) \
\sum_{a'}
\int \, d{\widehat z } \,  {\widehat z } \
C^{OUT}_{a^{\prime}\, j^{\prime}} \big({\widehat z } ,b\mu \big) \
e^{-S(b)}
 .
\label{eq:wtildedef}
\end{eqnarray}
The perturbative calculation of $\widetilde W(x,Q^2,b^2;j,j')$ is not
reliable for $b {\lower0.5ex\hbox{$\stackrel{>}{\sim}$}} 1/\Lambda$. However,
the integration over
$b$ in \eq{wdefii} runs to infinitely large $b^2$, and the
region $b {\lower0.5ex\hbox{$\stackrel{>}{\sim}$}} 1/\Lambda$ is important for
values of $Q^2$ and
${q_T^2}$ of practical interest. In order to deal with the large
$b^2$ region, we follow the method introduced in
Refs.~\cite{CS,CSS}. We define a value $b_{\rm max}$ such that we can
consider perturbation theory to be reliable for $b < b_{\rm max}$.
(In our numerical examples, we take $1/b_{\rm max} =2 \, {{\rm GeV}}$.)
Then we define a function $b_*$ of $b$ such that $b_* \approx b$ for
small $b$ and $b_* <  b_{\rm max}$ for all $b$:
\begin{equation}
b_* =
{ b \over \sqrt{1 + b^2/b_{\rm max}^2}}
\qquad .
\end{equation}

We define a version of $\widetilde W(x,Q^2,b^2;j,j')$ for which
perturbation theory is always reliable by $\widetilde
W(x,Q^2,b_*^2;j,j')$. Note that for small $b$, the difference between
$\widetilde W(b_*)$ and  $\widetilde W(b)$ is negligible because $b_* \approx
b$. Conversely, perturbation theory is always applicable for
the calculation of $\widetilde W(b_*)$ because $b_*$ is small even when
$b$ is large.

Next, we define a nonperturbative function $\exp(- S_{\rm NP}(b))$ as
the ratio of $\widetilde W(b)$ and $\widetilde W(b_*)$:
\begin{equation}
\widetilde W(x,Q^2,b^2;j,j')
=
\widetilde W(x,Q^2,b_*^2;j,j')\
e^{- S_{\rm NP}(x,Q^2,b^2;j,j') }
\qquad .
\label{eq:SNPdef}
\end{equation}
Ultimately, we will have to use nonperturbative information to
determine $S_{\rm NP}(b)$. However, some important information is
available to us. From \eq{wtildedef}, we see that
\begin{equation}
\dfrac{\partial \log[\widetilde W(x,Q^2,b^2;j,j')]}{\partial \log Q^2}
\end{equation}
is independent of $x,j,j'$ and $Q^2$. This result is derived in
perturbation theory, but at arbitrary order, so we presume that it
holds even beyond perturbation theory. Then
\begin{equation}
\dfrac{\partial S_{\rm NP}(x,Q^2,b^2;j,j')}{\partial \log Q^2 }
\end{equation}
is also independent of $x,j,j',k$ and $Q^2$. That is, $S_{\rm NP}$
has the form
\begin{equation}
S_{\rm NP}(x,Q^2,b^2;j,j') =
\log\left(Q^2/Q_0^2\right)g_1(b)
+ \Delta S_{\rm NP}(x,b^2;j,j')
\qquad .
\label{eq:sudakovi}
\end{equation}
(Here $Q_0$ is an arbitrary constant with dimensions of mass, inserted
to keep the argument of the logarithm dimensionless.) Furthermore, in
$\widetilde W$, the $x$ and $j$ dependence occurs in a separate factor from
the $j'$ dependence. Thus the second term in \eq{sudakovi} above
can be simplified to
\begin{equation}
S_{\rm NP}(x,Q^2,b^2;j,j') =
\log\left(Q^2/Q_0^2\right)g_1(b)
+ g_A(x,b^2;j)
+ g_B(b^2;j')
\qquad .
\end{equation}
(Recall, we have integrated over ${\widehat z }$.)

Perturbation theory is not applicable for the calculation of the
functions $g_1(b)$, $g_A(x,b^2;j)$ and $g_B(b^2;j')$  for large $b$.
For small $b$, perturbation theory tells us only that these functions
approach 0 as $b \to 0$. This follows from  \eq{SNPdef}, and
the fact that $b_*/b \to 0$ when $b \to 0$. (See Ref.~\cite{CSS} for
further discussion.) Since we learn little from perturbation theory,
we turn to non-perturbative sources of information. Fortunately, the
analogous functions in $e^+e^-$ annihilation and in the Drell-Yan
process have been fit using experimental
results.\citex{CS,DWS,yuanladinsky}

\figtempv   

We therefore ask whether the functions $g_1(b)$, $g_A(x,b^2;j)$ and
$g_B(b^2;j')$ in deeply inelastic scattering are related to the
analogous functions in the other two processes. Consider first
$g_1(b)$, the coefficient of $\log(Q^2/Q_0^2)$. According to the
analysis of Ref.~\cite{CS}, this function receives contributions from the
two jet subdiagrams in \xfig{nonpertiii}(b). (In this
figure, we use a space-like axial gauge.) The soft gluon connections
in \xfig{nonpertiii}(b) affect $g_A(x,b^2;j)$ and
$g_B(b^2;j')$, but do not contribute ``double logarithms,'' and thus
do not affect $g_1(b)$. Thus
\begin{equation}
g_1(b) \equiv g_1^{DIS}(b) = g_1^{\rm IN}(b) + g_1^{\rm OUT}(b)
\qquad ,
\end{equation}
where $g_1^{\rm IN}(b)$ is associated with the incoming beam jet
(the lower subdiagram in \xfig{nonpertiii}(b)) while
 $g_1^{\rm OUT}(b)$ is associated with the outgoing struck-quark jet
(the upper subdiagram in \xfig{nonpertiii}(b)). In the
Drell-Yan process, depicted in \xfig{nonpertiii}(a),
there are two incoming beam jets and one has
\begin{equation}
g_1^{\rm DY}(b) = 2 g_1^{\rm IN}(b)
\qquad .
\end{equation}
In $e^+ e^-$ annihilation, depicted in
\xfig{nonpertiii}(c), there are two outgoing quark
jets and one has
\begin{equation}
g_1^{e\bar e}(b) = 2 g_1^{\rm OUT}(b)
\qquad .
\end{equation}
Thus
\begin{equation}
g_1(b) \equiv g_1^{DIS}(b) =
 (1/2)\,g_1^{\rm DY}(b) +(1/2)\, g_1^{e\bar e}(b)
\qquad .
\end{equation}
In the following section, we show numerical results using
Ref.~\cite{CS} for $g_1^{e\bar e}(b)$ and
Ref.~\cite{DWS} for $g_1^{\rm DY}(b)$.

The situation for $g_A(x,b^2;j)$ and $g_B(b^2;j')$ is not so simple.
Let us write
\begin{equation}
S_{\rm NP}^{\rm DY}(x,Q^2,b^2;j,j') =
\log\left(Q^2/Q_0^2\right) g_1^{\rm DY}(b)
+ g_2^{\rm DY}(x_A,b^2;j)
+ g_2^{\rm DY}(x_B,b^2;j')
\qquad .
\end{equation}
for the Drell-Yan process and
\begin{equation}
S_{\rm NP}^{\bar e e}(x,Q^2,b^2;j,j') =
\log\left(Q^2/Q_0^2\right) g_1^{\bar e e}(b)
+ g_2^{\bar e e}(b^2;j)
+ g_2^{\bar e e}(b^2;j')
\qquad .
\end{equation}
for the energy-energy correlation function in $e^+e^-$ annihilation.
({\it Cf.}, \xfig{tempv}.)
One might like to assume that $g_A(x,b^2;j)$ is the same function as
$g_2^{\rm DY}(x,b^2;j)$ while $g_B(b^2;j')$ is the same function as
$g_2^{\bar e e}(b^2;j')$. However, this may not be true because all
of these functions get contributions from the soft gluon exchanges
that link the two jets in \xfig{nonpertiii},
(represented by  the function $U(b)$ in Ref.~\cite{CS}).
Furthermore, the
dependence of the functions $g_2^{\rm DY}(x,b^2;j)$ and
$g_2^{\bar e e}(b^2;j')$ on the flavors $j$ and $j'$ has not been
determined from experimental data. What we know are flavor averaged
functions $g_2^{\rm DY}(x,b^2)$ and $g_2^{\bar e e}(b^2)$. Thus the
best we can do is propose a model for the functions we need:
\begin{equation}
g_A(x,b^2;j) + g_B(b^2;j')
= t\, g_2^{\rm DY}(x,b^2)
+ (1-t)\,g_2^{\bar e e}(b^2)
\qquad ,
\label{eq:sudparm}
\end{equation}
where $0<t<1$, with $g_2^{\bar e e}(b^2)$ taken from
Ref.~\cite{CS} and $g_2^{\rm DY}(x,b^2)$ taken from
Ref.~\cite{DWS}. We vary the parameter $t$
between 0 and 1 to get an estimate of the uncertainty involved.
({\it Cf.}, \xfig{tempvi}.)

\figtempvi  
\figtempvia 

For comparison, we present the above parameterizations  for the
non-perturbative contributions with the recent fit by Ladinsky and
Yuan\citex{yuanladinsky} for W-production in \xfig{tempvia}. Ladinsky and
Yuan introduce an extra degree of freedom by  allowing for a $\tau=x_A\,
x_B$ dependence. We present the comparison for a range of $\tau$; this
allows one to gauge the effects of different non-perturbative
estimates, and correlate the Ladinsky and
Yuan parameterization with that presented in \eq{sudakovi} and
\eq{sudparm}.

\section{Reprise}

For the benefit of the reader, we review the principal steps
in the calculation of the energy distribution.  The  energy
distribution is given by:
\begin{eqnarray}
{ d \Sigma \over dx \, dQ^2 \, d{q_T^2}\, d\phi}
&=&
\sum_{k=1}^{9}  \
{ d \Sigma_k \over dx \, dQ^2 \, d{q_T^2}\, d\phi}
\nonumber \\[10pt]
&=&
\sum_{k=1}^{9}  \
{\cal A}_k(\psi,\phi)
\sum_{V_1,V_2}  \
\sum_{j, j'} \
\Sigma_0(Q^2;V_1,V_2,j, j',k) \
\Gamma_k(x,Q^2,q_T^2;j,j')
\qquad ,
\label{eq:}
\end{eqnarray}
where  ${\cal A}_k(\psi,\phi)$ are the nine angular
functions  arising from hyperbolic $D^1(\psi,\phi)$
rotation matrices. The sum on $V_1$
and $V_2$ runs over vector boson types, $\{\gamma,Z\}$ or
$\{W^\pm\}$ as appropriate.
 The sums over $j$ and $j^\prime$ include all quark
flavors, $\{u,\bar u,d,\bar d,\dots\}$; for neutral
currents, this sum is diagonal
$(j=j^\prime)$.
 The function
$\Sigma_0(Q^2;V_1,V_2,j, j',k)$  includes factors for the
coupling of the electron to the vector bosons as well as
factors for the propagation of the vector bosons.
 The energy distribution
function that we have computed is
$\Gamma_k(x,Q^2,q_T^2;j,j')$.

 In the limit ${q_T}\to 0$, the $\Gamma_1$ and $\Gamma_6$
will contain the dominant singularities as their angular
structure is proportional to the Born process.
 We define:
\begin{eqnarray}
\Gamma_1(x,Q^2,{q_T^2};j,j') &=&
 \Gamma_1^{Pert}(x,Q^2,{q_T^2};j,j')
\nonumber \\
&+&
  {\cal T}\left(\dfrac{{q_T}}{Q}\right)
\left\{ \phantom{(-1)} W^{}(x,Q^2,q_T^2;j,j') -
\Gamma_1^{Asym}(x,Q^2,{q_T^2};j,j') \right\}
 \nonumber\\
\Gamma_6(x,Q^2,{q_T^2};j,j') &=&
 \Gamma_6^{Pert}(x,Q^2,{q_T^2};j,j')
\nonumber \\
&+&
  {\cal T}\left(\dfrac{{q_T}}{Q}\right)
\left\{(-1) W^{}(x,Q^2,q_T^2;j,j') -  \Gamma_6^{Asym}(x,Q^2,{q_T^2};j,j')
\right\}
\label{eq:}
\end{eqnarray}
where the matching function ${\cal T}({q_T}/Q)$ [\eq{rhoeq}] is provided
to ensure proper behavior as ${q_T}\to Q$.
  $\Gamma_k^{Pert}$ represents the perturbative results of
paper~I [\eq{partonhadron}] calculated at order ${\alpha_s}^1$,
 $\Gamma_k^{Asym}$ represents the asymptotic limit
(${q_T}\to 0$) of
 $\Gamma_k^{Pert}$  [\eq{sudexp}], and
 $W^{}(x,Q^2,q_T^2;j,j')$ represents the summed
(Sudakov) term [\eq{suddefi}] which is finite as ${q_T}\to 0$.
 Note the function $W^{}(x,Q^2,q_T^2;j,j')$ the same for
both $\Gamma_1$ and $\Gamma_6$.

The form of the Sudakov structure function
is particularly simple in impact parameter space:
\begin{eqnarray}
W^{}(x,Q^2,q_T^2;j,j')
&=&
\int \, {d^2 b \over (2\pi)^2 } \
e^{i {q_T} \cdot b } \
\widetilde{W}^{}(x,Q^2,b^2;j,j')
\qquad .
\label{eq:}
\end{eqnarray}
 To ensure that the calculation is reliable for
large $b$ (small ${q_T}$), we introduce:
\begin{equation}
\widetilde W(x,Q^2,b^2;j,j')
=
\widetilde W(x,Q^2,b_*^2;j,j')\
e^{- S_{\rm NP}(x,Q^2,b^2;j,j') }
\qquad ,
\label{eq:}
\end{equation}
where $b_* \in [0,b_{max}]$ for $b \in [0,\infty]$.

The perturbative function $\widetilde W(x,Q^2,b_*^2;j,j')$
is given by:
\begin{eqnarray}
\widetilde W(x,Q^2,b_*^2;j,j')
&=&
\int_{x}^{1}
\dfrac{d\xi}{\xi} \
\sum_a \
f_{a/A}(\xi,\mu) \
C^{IN }_{ja} \big({\widehat x } ,b_*\mu \big) \
\int \, d{\widehat z } \,  {\widehat z } \ \sum_{a'} \
C^{OUT}_{a^{\prime}\, j^{\prime}} \big({\widehat z } ,b_*\mu \big) \
e^{-S(b_*)}
 ,
\nonumber \\
\label{eq:}
\end{eqnarray}
 where ${\widehat x }=x/\xi$.
 For the incoming particles, there is an integration over a parton momentum
fraction
$\xi$, a sum over parton types
$a = g, u, \bar u, d, \bar d, \dots$, a parton distribution
function $f_{a/A}$ and a set of perturbative coefficients
$C^{\rm IN }$.
  For the outgoing partons, there is an integration over
parton momentum fraction $\hat z$, weighted by $\hat z$,
a sum over parton types
$a' = g, u, \bar u, d, \bar d, \dots$, and there are
perturbative coefficients $C^{\rm OUT}$ associated  with
the outgoing states.
 The heart of the formula is the Sudakov factor
$\exp[-S(b_*)]$, defined as:
\begin{eqnarray}
S(b_*)
&=&
\int_{C_1^2/b_*^2}^{C_2^2 Q^2} \
{d\mu^2 \over \mu^2}
\left\{\ln\left[{C_2^2 Q^2\over \mu^2}\right]
A( \alpha_s(\mu) ) +
B( \alpha_s(\mu) )\right\}
\qquad .
\label{eq:}
\end{eqnarray}
The functions $A$, $B$, as well as  $C^{\rm IN}$ and
$C^{\rm OUT}$, have perturbative expansions in
powers of $\alpha_s$.
 We choose the arbitrary constants $\{ C_1, C_2\}$ as in \eq{c1c2}.

The non-perturbative
contribution is parameterized in terms of the fits to
$e^+ e^-$ and Drell-Yan data\rlap.\citex{eedata,CS,DWS}
\begin{equation}
S_{\rm NP} (x,Q^2,b^2;j,j') =
\log\left[\dfrac{Q^2}{Q_0^2}\right]
\left\{
\dfrac{g_1^{\rm DY}(b) +  g_1^{e\bar e}(b) }{2}
\right\}
+  t \  g_2^{\rm DY}(x,b^2)
+ (1-t) \ g_2^{\bar e e}(b^2)
\qquad .
\end{equation}
The arbitrary parameter $t \in [0,1]$
interpolates between the
$e^+ e^-$ and Drell-Yan form.


\section{Results}

We  present numerical results of the energy distribution function
for representative values of $\{x,Q^2\}$   using the CTEQ3 parton
distributions\rlap.\citex{cteq3}
 We present  results only for the $\Gamma_1$ set of structure functions,
as the $\Gamma_6$ set have the identical ${q_T}\to 0$ structure (up to a
sign).
 Recall that the structure functions are given by:
\begin{eqnarray}
\Gamma_1(x,Q^2,{q_T^2};j,j') &=&
 \Gamma_1^{Pert}(x,Q^2,{q_T^2};j,j')
\nonumber \\
&+&
  {\cal T}\left(\dfrac{{q_T}}{Q}\right)
\left\{ W^{}(x,Q^2,q_T^2;j,j') -  \Gamma_1^{Asym}(x,Q^2,{q_T^2};j,j') \right\}
\qquad .
\label{eq:}
\end{eqnarray}
Making use of \eq{master}, we have a parallel relation for the
energy distribution function:
\begin{eqnarray}
\dfrac{d\Sigma_1(x,Q^2,{q_T^2};j,j')}{dx\, dQ^2 \, d{q_T^2}\, d\phi} &=&
\dfrac{d\Sigma_1^{Pert}(x,Q^2,{q_T^2};j,j')}{dx\, dQ^2 \, d{q_T^2}\, d\phi}
\nonumber \\
&+&
 {\cal T}\left(\dfrac{{q_T}}{Q}\right)
\left\{
\dfrac{d\Sigma_1^{Sum}(x,Q^2,{q_T^2};j,j')}{dx\, dQ^2 \, d{q_T^2}\, d\phi}
- \dfrac{d\Sigma_1^{Asym}(x,Q^2,{q_T^2};j,j')}{dx\, dQ^2 \, d{q_T^2}\, d\phi}
 \right\}
 ,
\label{eq:}
\end{eqnarray}
 where we use the ``Sum" superscript to denote the summed Sudakov contribution
derived from $W$.
We will examine both the individual terms as well as the total in the
following.
 We will use the shorthand
$
d\Sigma_1 \equiv
d\Sigma_1(x,Q^2,{q_T^2};j,j')/(dx\, dQ^2 \, d{q_T^2}\, d\phi)
$

\subsection{${q_T}$ Distributions}

\figtempi   
\figtempiii 

In \xfig{tempi} and \xfig{tempiii}, we show  the separate contributions
to
$d\Sigma_1$ as a function of ${q_T}$
for
two choices of $\{x,Q^2\}$.\footnote{
In the small ${q_T}$ region,  $d\Sigma_1$ and $d\Sigma_6$ are independent of
$\phi$; therefore we need not specify it.
}
 We have included an extra factor of ${q_T^2}$ to make the features of the
plot more legible.
 As anticipated, we see that
$d\Sigma_1^{Pert}\simeq d\Sigma_1^{Asym}$ as ${q_T}\to 0$ leaving
$d\Sigma_1 \simeq d\Sigma_1^{Sum}$.
 For large ${q_T}$, we find  $d\Sigma_1^{Sum}- d\Sigma_1^{Asym} \simeq 0$,
but this cancellation is not as precise as the above because the relation
$\Gamma^{Sum} - \Gamma^{Asym} \simeq 0$ holds only to first-order.
 Therefore,   in the following figures
we shall include the ${\cal T}({q_T^2}/Q^2)$ factor to ensure that
$d\Sigma_1^{Sum}- d\Sigma_1^{Asym}$ is smoothly turned off at large
${q_T}$.
 The fact that $d\Sigma_1^{Sum}$ and $d\Sigma_1^{Asym}$
become negative for large ${q_T}$ reminds us that these expressions
were  approximations valid only for ${q_T} \ll Q$.

\figtempii  
\figtempiv  

Having examined the separate terms, we now turn our attention to
the  energy distribution  function,
$d\Sigma_1$.
 Again, we have included an extra factor of ${q_T^2}$
 in \xfig{tempii}(a) and \xfig{tempiv}(a)  to make the features of the
plot more legible.
 In \xfig{tempii}(b) and
\xfig{tempiv}(b), we plot $d\Sigma_1$ in
the small ${q_T}$ region (without an extra ${q_T^2}$  factor) to demonstrate
that the summed results approach a finite limit as ${q_T}\to 0$.
 We present the results for three choices of the non-perturbative
function $S_{\rm NP}(x,Q^2,b^2;j,j')$ as parameterized in \eq{sudparm}.
 The choice $t=0$ corresponds to the $e^+e^-$ limit\rlap,\citex{CS}
 while $t=1$ corresponds to the Drell-Yan limit\rlap,\citex{DWS}
 and $t=1/2$ corresponds to an even mix of the above.
 The difference due to the non-perturbative contribution is quite
significant for low ${q_T}$.
 The  $t=0$  ($e^+e^-$)
non-perturbative  function, which is much narrower in $b$-space, yields a
broader energy  distribution; this is clearly evident in the
figures as we see the peak move to lower ${q_T}$ values as we shift from
the  $t=0$  ($e^+e^-$)
to   $t=1$  (Drell-Yan).
 At large ${q_T}$, $d\Sigma_1$  is independent
of the non-perturbative contributions, since it is dominated by
$d\Sigma_1^{Pert}$.

Clearly, the HERA data should be able to distinguish between this range
of distributions, particularly in the small ${q_T}$ regime where the
span of the   non-perturbative contributions are
significant.\citex{wsmith,olnesstung}

\section{Conclusions}

Measurement of the distribution of hadronic energy in the final state in
deeply inelastic electron scattering at HERA can provide a good test of
our understanding of perturbative QCD.
 Furthermore, we can probe non-perturbative physics because the
the energy distribution functions are sensitive to  the  non-perturbative
Sudakov form factor $S_{{{\scriptscriptstyle N\kern-0.16667em P}}}(b)$ in the
small ${q_T}$ region.

 We have evaluated the energy distribution function
for finite transverse momentum ${q_T}$ at order
$\alpha_s$ in paper~I.
 Because the distribution is weighted  by the final state hadron energy,
this physical observable is  infrared safe, and independent of the
decay distribution functions.
  In this paper, we sum the soft gluon radiation into a Sudakov form
factor to  evaluate the  energy distribution function in the small
${q_T}$ limit.
 By matching the small and large ${q_T}$ regions, we  obtain a complete
description throughout the kinematic range.
 This result is significant phenomenologically as a the bulk of the events
occur at small ${q_T}$ values, where perturbation theory by itself is
divergent.  This technique can provide an incisive tool for the
study of deeply inelastic scattering.
 Additionally, crossing relations allow us to relate the
non-perturbative contribution in
deeply inelastic
scattering   energy distributions to analogous quantities in the Drell-Yan
and $e^+e^-$ annihilation processes.

\acknowledgments

We would like to thank
E. Berger,
S. Ellis,
K. Meier,
W. Tung,
for valuable discussions.
 We also thank  R. Mertig for assistance with FeynCalc, and
S. Riemersma for carefully reading the manuscript.
R.M and F.O.  would also like to acknowledge the support and  gracious
hospitality of  Dr. A. Ali and the
Deutsches Elektronen Synchrotron.
This work is supported in part by
the U.S. Department of Energy, Division of High Energy Physics.

\appendix

\section{Kinematic Relations}

We present some basic kinematic relations to facilitate the
calculation.
First we give the expressions to relate $\{E^{\prime},\theta^{\prime}\}$
to $\{x,Q^2\}$,
 \begin{eqnarray}
Q^2 &=& -q \cdot q =  2 E E^{\prime} ( 1 - \cos\theta^{\prime})
\\[10pt]
x &=&
\dfrac{Q^2}{2 q\cdot P_A}
=
  \dfrac{ E E^{\prime}  ( 1 - \cos\theta^{\prime}) }
   { E_A[ 2 E - E^{\prime}( 1 + \cos\theta^{\prime})]}
\qquad .
 \end{eqnarray}
Next, we give the expression for the Born scattering angle $\theta_*$,
 \begin{eqnarray}
\cot \left( \dfrac{\theta_*}{2} \right)
&=&
{ 2x E_A \over Q}
\left[1 - {Q^2 \over x s}\right]^{1/2}
\qquad .
\end{eqnarray}
 The corresponding azimuthal angle, $\phi_*$ is trivial, and can be
defined to be zero.
 Finally, we give the expressions to compute the natural variables
of the Breit frame, $\{{q_T},\phi\}$:
\begin{eqnarray}
{q_T^2}
&=&
{8E^2 - 4E'( 2E - E') (1+ \cos\theta^{\prime}  )\over 1 - \cos\theta_B} \
\left\{ \sin^2\left[\dfrac{\theta_B - \theta_*}{2}\right]
  +\sin\theta_B \sin\theta_* \
    \sin^2\left[\dfrac{\phi_B - \phi_*}{2} \right] \right\}
\nonumber \\
\end{eqnarray}
\begin{eqnarray}
\cos(\phi) &=&
{Q \over 2 {q_T}}
\left[
1 - {Q^2 \over x s}
\right]^{-1/2}\
\left\{
1
- {Q^2 \over x s}
+ {{q_T^2} \over Q^2}
-\left({Q \over 2x E_A}\right)^2
\cot \left( \dfrac{\theta_B}{2} \right)
\right\}
\qquad .
\end{eqnarray}

\section{Energy Distribution Formulas}

We now give some explicit formulas for computation of the
structure functions
and energy distribution contributions.
 The process we consider is the hadronic process
$e^- + A \to e^- + B + X$, and
the fundamental formula for computation of the
structure functions and energy distribution contributions is:
\begin{eqnarray}
{ d \Sigma \over dx \, dQ^2 \, d{q_T^2}\, d\phi}
&=&
\sum_{k=1}^{9}  \
{\cal A}_k(\psi,\phi)
\sum_{V_1,V_2}  \
\sum_{j, j'} \
\Sigma_0(Q^2;V_1,V_2,j, j',k) \
\Gamma_k(x,Q^2,q_T^2;j,j')
\qquad .
\label{eq:masterii}
\end{eqnarray}
with
\begin{eqnarray}
\Sigma_0(Q^2;V_1,V_2,j, j',k)
&=&
\dfrac{Q^6}{2^6 \pi  x^3 s^2 E_A} \
\dfrac{G_k^q(V_1,V_2;j, j') \
 G_k^\ell(V_1,V_2) }
{(Q^2 + M_{V_1}^2) (Q^2 + M_{V_2}^2)}
\qquad .
\label{eq:sigzero}
\end{eqnarray}
${\cal A}_k(\psi,\phi)$ represents the nine angular functions
arising from the  hyperbolic ${}^1D(\psi,\phi)$ rotation matrices.
$G_k^q(V_1,V_2;j, j')$ and
$G_k^\ell(V_1,V_2)$ are the combinations of couplings
from the leptonic and hadronic tensors, respectively, as  defined in paper~I.
$(Q^2 + M_{V_i}^2)$ arise from the boson propagators,
and
$\Gamma_k(x,Q^2,q_T^2;j,j')$ are the hadronic energy distribution
function.
 We sum over the intermediate vector bosons $\{V_1, V_2\}=
\{\gamma,Z^0\}$ or $\{W^\pm\}$, as appropriate, and the parton species
$\{ j, j' \}$.
\begin{eqnarray}
\begin{array}{rcll}
{\cal A}_{1}\, (\psi,\phi)  &=& (+1)                  & [1+{{\rm
cosh}}^2(\psi)]
\nonumber\\
{\cal A}_{2}\, (\psi,\phi)  &=&  (-2) &
\nonumber\\
{\cal A}_{3}\, (\psi,\phi)  &=& (-1) \ \cos(\phi )\   &{{\rm sinh}}(2\psi )
\nonumber\\
{\cal A}_{4}\, (\psi,\phi)  &=& (+1) \ \cos(2 \phi )\ &{{\rm sinh}}^2 (\psi )
\nonumber\\
{\cal A}_{5}\, (\psi,\phi)  &=& (+2) \ \sin(\phi )\   &{{\rm sinh}}(\psi )
\label{eq:As}\\
{\cal A}_{6}\, (\psi,\phi)  &=& (+2) & {{\rm cosh}}(\psi )
\nonumber\\
{\cal A}_{7}\, (\psi,\phi)  &=& (-2) \ \cos(\phi )\   &{{\rm sinh}}(\psi )
\nonumber\\
{\cal A}_{8}\, (\psi,\phi)  &=& (-1) \ \sin(\phi )\   &{{\rm sinh}}(2 \psi )
\nonumber\\
{\cal A}_{9}\, (\psi,\phi)  &=& (+1) \ \sin(2 \phi )\ &{{\rm sinh}}^2 (\psi )
\qquad .
\end{array}
\end{eqnarray}
\goodbreak
Note, for instance, the analogy between the angular coefficient
${\cal A}_{1}= 1+{{\rm cosh}}^2(\psi)$,
which appears in the order $\alpha_s^0$ energy distribution, and the
corresponding coefficient in the case of the Drell-Yan energy
correlation,
\hbox{$1 + \cos^2(\theta)$.\citex{angulartheory}}

\section{Davies, Webber, \& Stirling Parametrization}

The form of the non-perturbative Sudakov function $S_{{{\scriptscriptstyle
N\kern-0.16667em P}}}(b)$, used by
Davies, Webber, and Stirling to introduce
the transverse momentum smearing in the Drell-Yan process is:
\begin{eqnarray}
S_{{{\scriptscriptstyle N\kern-0.16667em P}}} (b)
&=&
b^2
\left[
g_1 + g_2 \ \ln \left( \dfrac{b_{max} Q}{2} \right)
\right]
\end{eqnarray}
with
\begin{eqnarray}
g_1 &=& 0.15 \, {{\rm GeV}}^2      \\
g_2 &=& 0.40 \, {{\rm GeV}}^2          \\
b_{max} &=& (2\,  {{\rm GeV}})^{-1}
\qquad .
\end{eqnarray}

\section{Collins \& Soper Parametrization}

The form of the non-perturbative function used by
Collins and Soper to introduce
the transverse momentum smearing in the $e^+\, e^-$ process is:
\begin{eqnarray}
S_{{{\scriptscriptstyle N\kern-0.16667em P}}} (b)
&=&
A \left\{
4 A_1 \ {\alpha_s(\mu) \over \pi} \
\ln\left[ C_2 \, Q \, b_{max} \over C_1  \right] \
\ln\left( {b \over b_*}   \right)
 \right\}
+ \Delta f_1(b) \  \ln\left({Q^2\over Q_0^2} \right)
+ \Delta f_2(b)
\end{eqnarray}
with
\begin{eqnarray}
\Delta f_1(b) &=&  A_{11} \, b + A_{12} \, b^2   \nonumber\\
\Delta f_2(b) &=&  A_{21} \, b + A_{22} \, b^2
\qquad .
\end{eqnarray}
While the functional form allowed here is quite general,
in practice,  it was possible to obtain a good fit to the data
using only the $A$ and $A_{21}$ parameters.
Specifically,
\begin{eqnarray}
A      &=& 1.33  \nonumber\\
A_{21} &=& 1.5   \nonumber\\
A_{11} &=&  A_{12} =  A_{22} =  0
\qquad .
\end{eqnarray}
Additional parameters and relations necessary are:
\begin{eqnarray}
b_{max}&=& (2\, {{\rm GeV}})^{-1}    \nonumber\\
Q_0 &=& 27 \, {{\rm GeV}}  \nonumber\\
\mu &=& C_1/b_* \nonumber\\
A_1 &=& 2 \, C_F
\qquad .
\end{eqnarray}

\section{Ladinsky \& Yuan Parametrization}

The form of the non-perturbative Sudakov function $S_{{{\scriptscriptstyle
N\kern-0.16667em P}}}(b)$, used by
Ladinsky and Yuan to introduce
the transverse momentum smearing in the Drell-Yan process is:
\begin{eqnarray}
S_{{{\scriptscriptstyle N\kern-0.16667em P}}} (b)
&=&
\left[
g_1 \, b^2 + g_1 \, g_3 \, b \ln[100 \, \tau]
+ g_2 \, b^2 \ \ln \left( \dfrac{Q}{2 \, Q_0}
\right)
\right]
\end{eqnarray}
with
\begin{eqnarray}
g_1 &=& 0.11 \, {{\rm GeV}}^2      \nonumber\\
g_2 &=& 0.58 \, {{\rm GeV}}^2          \nonumber\\
g_3 &=& -1.5 \, {{\rm GeV}}^{-1}  \nonumber\\
Q_0 &=& 1.60 \, {{\rm GeV}}^2          \nonumber\\
b_{max} &=& (2\,  {{\rm GeV}})^{-1}
\qquad .
\end{eqnarray}

\section{$\alpha_s$ at 1-Loop and 2-Loop}

To properly compute the $\mu^2$ integral in the Sudakov form factor,
it will be necessary to use the complete result for the running
coupling at both 1- and 2-loops.
The 2-loop result for $\alpha_s$ is:
\begin{eqnarray}
\alpha_s(\mu^2) &=&
{ 4 \pi  \over \beta_1 \ln(\mu^2/\Lambda^2) }
- { 4 \pi \beta_2 \, \ln[\, \ln(\mu^2/\Lambda^2)\, ] \over
\beta_1^3 \, \ln^2(\mu^2/\Lambda^2) }
\end{eqnarray}
where
\begin{eqnarray}
\beta_1 &=&  {(11N_c - 2 N_f) \over 3}  \equiv {(33-2 N_f) \over 3}  \\
\beta_2 &=&  (102   - {38N_f \over 3})
\qquad .
\end{eqnarray}
The 1-loop result is simply obtained by taking $\beta_2\to0$.

\section{Integral Table  \label{APPA}}

For simplicity and completeness, we list the integrals we shall encounter
in the Sudakov form factor at the 1- and 2-loop level.
We consider the logarithmic terms ($A_i$) and the constant terms
($B_i$) using the 2-loop expression for $\alpha_s$;
the 1-loop expressions are easily re covered in the limit $\beta_2 \to 0$.
 It will be convenient to define the following quantities:
\begin{eqnarray}
L_1 &=& \ln\left[{C_1^2 \over b^2  \Lambda^2}\right]
\nonumber \\[5pt]
L_2 &=& \ln\left[{C_1^2 \over b^2 C_2^2 Q^2}\right]
\quad \equiv \quad L_1 - L_3
 \\[5pt]
L_3 &=& \ln\left[{C_2^2 Q^2 \over \Lambda^2}\right]
\qquad .
\nonumber
\end{eqnarray}

First, the $A_1$ term with the 2-loop expression for $\alpha_s$.
\begin{eqnarray}
&&\int_{C_1^2/b_*^2}^{C_2^2 Q^2} \
{d\mu^2 \over \mu^2}
\ln\left[{C_2^2 Q^2\over \mu^2}\right]
{\alpha_s(\mu;2) \over (2) \pi} \,  A_1
=
{4 A_1 \over  (2)  \beta_1}  \
 \left(L_2  + L_3  \ln\left[{L_3 \over L_1} \right] \right)
\nonumber \\[5pt]  && \qquad
+{4 A_1 \beta_2 \over  (2)  \beta_1^3 }  \left\{
+ {L_2 \over  L_1 }
- {L_3  \ln[L_1 ] \over L_1 }
+ \ln[L_3 ]
+ {\ln[L_3 ]^2 - \ln[L_1 ]^2 \over 2 }
\right\}
\qquad .
\end{eqnarray}
The $B_1$ term with the 2-loop expression for $\alpha_s$.
\begin{eqnarray}
\int_{C_1^2/b_*^2}^{C_2^2 Q^2} \
{d\mu^2 \over \mu^2}
{\alpha_s(\mu;2) \over (2) \pi} \,  B_1
&=&
{4 B_1   \over (2)  \beta_1 } \ \ln\left[{L_3 \over L_1} \right]
\nonumber \\[5pt]
&+& {4 \beta_2 B_1   \over (2)  \beta_1^3 L_1 L_2 }
\left(
L_1 - L_3 + L_1 \ln[L_3]  - L_3 \ln[L_1]
 \right)
\qquad .
\end{eqnarray}
The $A_2$ term with the 1-loop expression for $\alpha_s$.
\begin{eqnarray}
\int_{C_1^2/b_*^2}^{C_2^2 Q^2} \
{d\mu^2 \over \mu^2}
\left({\alpha_s(\mu;1) \over (2) \pi} \right)^2 \,  A_2
&=&
{16 A_2 \over (4) \beta_1^2 L_1 }\
\left( -L_2 - L_1 \ln\left[{L_3 \over L_1} \right] \right)
\qquad .
\end{eqnarray}

\section{Boson-fermion couplings}

\goodbreak
%
\bigskip
\bigskip
\begin{center}
\nobreak
\begin{tabular}{||c||c|c||c|c||}\hline\hline

Fermions & $g_v(\gamma)$ & $g_a(\gamma)$
         & $g_v(Z)$ & $g_a(Z)$
\\ \hline\hline
{\vrule height 20pt depth 20pt width 0pt }$e^-$ & $-e$ & $0$ &
$-e\
{ 1-4 \sin^2\theta_W  \over  4\cos\theta_W \sin\theta_W} $ &
$+e\ {1\over  4\cos\theta_W \sin\theta_W  }  $
\\ \hline
{\vrule height 20pt depth 20pt width 0pt }$u,c,t$
& ${2 \over 3}\, e $ & $0$  &
$+ e\
{1-{8\over 3} \sin^2\theta_W \over  4\cos\theta_W \sin\theta_W}$ &
$-e\ {1 \over  4\cos\theta_W \sin\theta_W }  $
\\ \hline
{\vrule height 20pt depth 20pt width 0pt }$d,s,b$
& $-\,{1 \over 3}\, e$ & $0$ &
$-e\
{1-{4\over 3} \sin^2\theta_W \over  4\cos\theta_W \sin\theta_W}$ &
$+e\ { 1\over  4\cos\theta_W \sin\theta_W }  $
\\ \hline\hline
\end{tabular}
\nobreak
\end{center}
\nobreak
\medskip
\nobreak
\centerline{Table 1.  Boson-fermion couplings.}
\goodbreak


\end{document}